\documentclass[11pt]{article}

\usepackage{setspace}
\usepackage{graphicx} 
\usepackage{subfigure}
\usepackage{amssymb}
\usepackage{amsmath}
\usepackage{amsfonts}
\usepackage{psfrag}
\usepackage{url}      
\usepackage{boxedminipage} 

\newcommand{\prob}[1]{\mathsf{\Pr}\left(#1\right)}
\newcommand{\EXP}[1]{\mathsf{E}\left(#1\right)}

\newcommand{\remove}[1]{}

\begin{document}

\title{Estimating Network Link Characteristics using Packet-Pair
  Dispersion: A Discrete Time Queueing Theoretic View}

\author{
  \begin{tabular}{cc}
    Bikash Kumar Dey and D.~Manjunath & Supriyo Chakraborty\\
    Deptt. of Elecl Engg & Deptt. of Elecl. Engg. \\
    Indian Institute of Technology Bombay & University of California \\
    Powai Mumbai 400 076 INDIA & Los Angeles CA USa\\
    \texttt{bikash,dmanju,@ee.iitb.ac.in}  &
    \texttt{supriyo.chakraborty@gmail.com} 
  \end{tabular}
}

\maketitle
\begin{abstract} 
  Packet-dispersion based measurement tools insert pairs of probe
  packets with a known separation into the network for transmission
  over a unicast path or a multicast tree. Samples of the separation
  between the probe pairs at the destination(s) are observed.
  Heuristic techniques are then used by these tools to estimate the
  path characteristics from the observations. In this paper we present
  a queueing theoretic setting for packet-dispersion based probing.
  Analogous to network tomography, we develop techniques to estimate
  the parameters of the arrival process to the individual links from
  the samples of the output separations, i.e., from the end-to-end
  measurements.

  The links are modeled as independent discrete time queues with
  i.i.d.~arrivals. We first obtain an algorithm to obtain the (joint)
  distribution of the separation between the probes at the
  destination(s) for a given distribution of the spacing at the input.
  The parameter estimates of the arrival process are obtained as the
  minimizer of a cost function between the empirical and calculated
  distributions. We also carry out extensive simulations and numerical
  experiments to study the performance of the estimation algorithm
  under the fairly `harsh' conditions of non stationarity of the
  arrival process. We find that the estimations work fairly well for
  two queues in series and for multicast.

 \end{abstract}
 

\section{Introduction and Background}
\label{sec:introduction}

Network tomography is the estimation of detailed statistics of
performance parameters from aggregate or end-to-end measurements of
some measurable quantity. The term was first coined in \cite{Vardi96},
where the problem of estimating flow volumes from link volumes was
analyzed.  This is also called the traffic matrix estimation problem
and has been extensively studied in the transportation literature, for
which, an excellent survey is available in \cite{Abramsson98}.  The
traffic estimation problem is now being addressed by the networking
community
(\cite{Zhang03a,Zhang03b,Medina03,Lakhina04,Papagiannaki04}. Another
network tomography problem is that of estimating link delay statistics
from the samples of end-to-end delays experienced by multicast packets
\cite{Coates01,Tsang01}. Much of the work in this area has been in
estimating the link or segment delay distributions from the end-to-end
measurements. See \cite{Castro04} for a fairly comprehensive survey of
both delay and traffic estimation tomography problems and solutions
related to networking.

The bandwidth on a network path has also been of significant interest
and a number of tools have been developed for its estimation. Two
kinds of bandwidth related metrics are often estimated. The
\emph{bottleneck bandwidth} is the maximum transfer rate that could be
achieved over a path and is essentially the minimum of the
transmission rates on the links in the path. The \emph{available
  bandwidth} is the portion of the bottleneck bandwidth not used by
competing traffic and depends on the traffic load at the inputs to the
links.

Many tools have been developed to measure the bottleneck and available
bandwidths. To the best of our knowledge, all these tools are based on
either `packet-pair' or `packet-dispersion' techniques.  The
packet-pair technique was first described in \cite{Keshav91} to
measure the bottleneck bandwidth on a path where the links are
rate-servers. In this technique two back-to-back packets of equal
length are transmitted by the source.  It can be shown that the ratio
of the probe packet length to the separation between them at the
receiver is the service rate for the packet at the bottleneck link on
the path.  This idea has been used to develop many tools, e.g.,
\texttt{pathchar} \cite{Jacobson97} and \texttt{clink}
\cite{Downey99}, that measure the bottleneck capacity of a network
path. Inspired by the packet-pair technique, the packet-dispersion,
also called packet-spacing, technique has been developed to measure
the available bandwidth of an Internet path
\cite{Jain02,Rebeiro03,Prasad03,Strauss03,Chakraborty05}. In this
technique, a source transmits a number of probe packets with a
predetermined separation, say $d.$ The samples of the separation at
the receiver, say $d_r,$ are then used to estimate the available
bandwidth on the path.  \texttt{pathload} \cite{Jain02} and
\texttt{pathChirp} \cite{Rebeiro03} are some examples of tools that
use the packet-spacing technique to measure the available bandwidth.
See \cite{Prasad03} for an excellent survey of the different
packet-spacing techniques used in the measurement tools and
\cite{Strauss03} for an experimental comparison of the tools. A non
cooperative version that does not require a measurement-enabled
receiver is described in \cite{Chakraborty05}.

Packet-pair and packet-dispersion techniques have an important
advantage in that they do not require the sender and the receiver to
be synchronized in time. This makes them important in practice.
However, the tools that employ these techniques are based on informal
arguments. There have been excellent statistical studies based on
experiments on packet-pair techniques (e.g.,
\cite{Dovrolis01,Pasztor03}). Recently, there have been a few
theoretical models developed for packet dispersion based bandwidth
probing techniques
\cite{Liu05,Liu07,Liu08,Liebeherr07,Park06,Haga06,Machiraju07}. In
\cite{Liu05,Liu07} it is shown that the output dispersion of probe
pair samples the sum of three correlated random processes that are
derived from the traffic arrival process. The asymptotics of these and
the probing process are also analysed. The measurement methods of some
bandwidth estimation tools are then related to the analytical models
developed. These results are then extended to multihop paths in
\cite{Liu08}. In \cite{Liebeherr07}, the network is assumed to be a
time-invariant, min-plus system that has an unknown service
curve. This service curve estimation method is developed for different
probing techniques. 

While the above works provide interesting theoretical insights into
the probing process, our work is more closely related to that of
\cite{Machiraju07,Park06,Haga06}. In \cite{Machiraju07}, a probe train
is assumed and the actual delay of each of the probe packets is
assumed to be known at the output of the path. The delay sequence is
assumed to form a Markov chain whose transition probability matrix is
estimated. An inversion is defined to estimate the characteristics of
the arrival process from the transition probabilities of the delay
Markov chain. In \cite{Park06}, the bottleneck link is modeled as an
M/D/1 queue, and the expected dispersion is calculated using the
transient analysis. The method of moments is then used to estimate the
arrival rate. A similar method is followed in \cite{Haga06} where the
more general M/G/1 queue is considered and Takacs'
integro-differential equation is used to obtain the distribution of
the dispersion.

In this paper we develop techniques to estimate the parameters of the
arrival process distribution on the individual links on a path or a
multicast tree from packet-dispersion samples in a queuing theoretic
setting. Packet-pair probes with predetermined separation between them
are injected at a source.  These probe packets traverse discrete-time
queues with i.i.d.~packet arrivals on a path or in a multicast tree.
Samples of the separation between the probe packets at the output
nodes of the path or the multicast tree are used to estimate the
parameters of the arrival process distribution at the individual
links. The approach is as follows. First, for a single queue and for a
given separation between the probes at the input, we derive the
conditional distribution of the separation between the probes at the
output of the queue in terms of the distribution of the arrival
process to the discrete time queue.  The key intermediate result here
is the joint distribution of the number arrivals to the queue and the
number of departures from the queue between the slots in which the
probes are injected. Then the output separation is obtained for any
given distribution of the input separation.  Since the input
separation distribution is known, (we assume it to be fixed) this is
applied recursively on the path to obtain the separation distribution
at the outputs. The parameter estimator is the minimizer of a suitable
distance function between the empirical and the theoretical
distributions of the output separations.

We first develop our results by assuming that the probe packets are
served after all the packets that arrive in their slot, i.e., the
probe packets have the least priority among the packets that arrive in
their slot. Extension to the case of the probe packet having the same
priority (or any arbitrary fixed priority) as the other packets that
arrive in the slot will be along the same lines and is also presented.
We present numerical results only for the first case, i.e., when the
probes have lowest priority.

Our modeling assumption is similar to that of \cite{Machiraju07} in
that we also consider discrete time queues. While \cite{Machiraju07}
considers the delay process of the probe sequence, we consider the
dispersion of the packet pair. The inversion methods are also
different. While we consider a general i.i.d. arrival process for the
discrete time queeu, \cite{Park06} considers an M/D/1 queue and
obtains approximations, \cite{Haga06} considers an M/G/1 queue. 

The rest of the paper is organized as follows. In the next section, we
describe the problem setting and the notation that we will be using in
the paper. In Section~\ref{sec:output-dispersion} we consider a single
pair of probes that are injected into a stationary discrete-time queue
and develop an algorithm for computing the distribution of the
separation between a pair of probes at the output of a single queue
when the distribution of the separation at the input of the queue and
also the arrival process distribution are known. In
Section~\ref{sec:probe-pair-network}, we consider sending the probes
over a multi-link path and over a multicast tree and obtain the
distribution of the output separation(s) between the probes at their
outputs.  In Section~\ref{sec:estimating-A_n} we describe the method
to estimate the parameters of the arrival process at the queues. The
numerical evaluation of the estimates for the case of Poisson arrivals
are discussed in detail in Section~\ref{sec:numerical-results}. In
Section~\ref{sec:generalizations}, we describe a generalization and an
extension to the basic theory developed in
Section~\ref{sec:output-dispersion}. In particular, we outline the
method for computing the output separation distribution when the
probes have equal priority with the packets arriving in the same slot.
We conclude with a discussion in Section~\ref{sec:discussion} where we
explore connections with an earlier theoretical study of
packet-dispersion techniques. We also contrast our results with
delay-tomography results.

\section{Notation and Preliminaries}
\label{sec:notation}
\begin{figure}
  \begin{center}
    \     
    \includegraphics[width=4.5in]{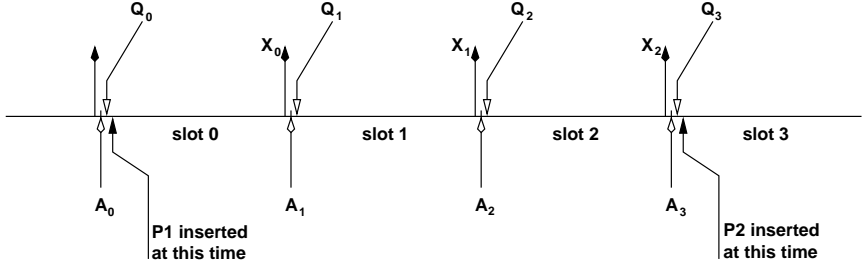}
  \end{center}

  \caption{The convention for $A_n,$ $X_n$ and $Q_n.$}
    \label{fig:convention}
\end{figure}

We first describe the notation used in this paper. For pedagogical
purposes, we consider a packet queuing system. In this section we will
consider a single-server queue with infinite buffers and FCFS service
discipline.  The service time of each packet is equal to the slot
length. The convention will be as follows.  Packet arrivals to the
queue in slot $n$ will occur at the beginning of the slot and the
departures will occur at the end of the slot. Thus arrivals in a slot
will be available for departure in the slot. We observe the queue at
the beginning of a slot, just after the arrival instant. Let $A_n$
denote the number of packets that arrive in slot $n,$ $X_n$ the number
of departures in slot $n$ and $Q_n$ the number of packets in the queue
when the queue is observed at the beginning of the slot. This
convention is shown in Figure~\ref{fig:convention}. Of course, $X_n$
can only be $0$ or $1.$ Further, for $m \leq n,$ $A_{m,n}:=A_m +
A_{m+1} + \cdots + A_n$ will denote the cumulative arrivals in the
interval $[m,n]$ and $X_{m,n}:=X_m + X_{m+1} + \cdots + X_n$ will
denote the cumulative departures in the interval $[m,n].$ In
particular, $X_{m,m}=X_m$ and $A_{m,m}=A_m.$ Note that $X_{m,n}$ is a
function of $Q_m$ and $[A_m,\ldots, A_n].$ $X^q_{m,n}$ will denote the
number of departures in slots $[m,n]$ that the same arrival sequence
$[A_m,\ldots, A_n]$ would cause with $Q_m=q.$ Observe that this makes
$X^q_{m,n}$ independent of the distribution of $Q_m$ and a function of
only $[A_m,\ldots, A_n].$ The work-conserving, FCFS service discipline
will be applied at the granularity of the batch of packets that arrive
in a slot and the packets in a batch will be queued in a random order.
The queue evolution equation as per the above convention will be
\begin{eqnarray}
  Q_{n+1} &=& Q_n + A_{n+1} - X_{n}, \nonumber \\
  Q_{n+m} &=& Q_n + A_{n+1,n+m} - X_{n,n+m-1}
  \label{eq:queue-evolution}
\end{eqnarray}

In the rest of this paper we assume that the arrival sequence $A_n$ is
i.i.d.~and let $p_k:=\prob{A_0=k}.$ Recall that if $\lambda:=\EXP{A_0}
< 1$ then the queue length process is stationary and the moment
generating function $\mathcal{Q}(z)$ of the stationary queue length
distribution $\pi_q:=\prob{Q_0=q}$ is given by
\begin{displaymath}
  \mathcal{Q}(z) =
  \frac{(1-\lambda)(z-1)\mathcal{P}(z)}{z-\mathcal{P}(z)} . 
\end{displaymath}
where $\mathcal{P}(z)$ is the moment generating function of $p_k.$  In
fact $\pi_q$ can be obtained using the following recursion.
\begin{eqnarray}
  \pi_q & := & \prob{Q_0=q} \nonumber \\
  &=& 
  \begin{cases} 
    1 - \EXP{A_0}  \ = \ 1 - \lambda & \mbox{for $q=0$,} \\
    \frac{1}{p_0}\left( \pi_0(1-p_0) \right) & \mbox{for $q=1$,} \\
    \frac{1}{p_0}\left( \pi_{q-1}(1-p_1) - \pi_{q-2} p_2 -
      \cdots \right. & \\
    \hspace{0.5in} \left. - \ \pi_1 p_{q-1} - \pi_0 p_{q-1}  \right) &
    \mbox{for $q > 1$.} 
  \end{cases} \nonumber \\
  \label{eq:pi_q-recursion}  
\end{eqnarray}

\section{Passage of a Probe Pair through a Discrete Time Queue}
\label{sec:output-dispersion}
In this section we consider injecting a single probe pair into a
stationary queue and derive the distribution of the output separation,
$d_o,$ as a function of the input separation, $d_i,$ and the
parameters of the packet arrival process $\{A_n\}.$ For a given
distribution of the input separation, the distribution of $d_o$ is
obtained as
\begin{equation}
  \label{eq:d_o-unconditional}
  \prob{d_o=s} =\sum_{j=1}^\infty \prob{d_o=s|d_i=j} \prob{d_i=j} . 
\end{equation}

Let us consider a queue in steady state. Consider two probe packets,
denoted by $P_1$ and $P_2,$ inserted into the discrete time queue in
slots $0$ and $d_i$ respectively. Since we are considering stationary
queues, without loss of generality, we can relabel the slots in this
manner. These probe packets are enqueued along with the other packets
that arrive in their respective slots and are available for
transmission in the same slot in which they arrive. In this section,
we assume that the probe packets have the least priority among the
packets that arrive in their slot.

Let $D_1$ and $D_2$ denote the departure slots for the first and the
second probe respectively. Let $d_o := D_2 - D_1$ denote the
dispersion of the probe packets at the output of the queue.  Clearly,
\begin{eqnarray}
  D_1 &=& Q_0  \ , \nonumber \\
  D_2 &=& d_i + Q_{d_i} \ = \ d_i + Q_0 + 1 + A_{1,d_i} - X_{0,d_i-1}
  \ , \nonumber \\  
  d_o := D_2 - D_1 &=& d_i + 1 + A_{1,d_i} - X_{0,d_i-1}  \ .
  \label{eq:one-queue-dispersion}
\end{eqnarray}
$d_o-d_i$ is the \emph{packet-dispersion}. In the above, $Q_0$ does
not include probe packet $P_1$ and $Q_{d_i}$ is the number of packets
in the queue at the beginning of slot $d_i$ not including $P_2.$ Also,
$A_{d_i}$ does not include $P_2$ while $X_{0,d_i-1}$ includes the
possible departure of $P_1.$ Note that $X_{0,d_i-1} \geq X_0 =1$ since
the queue has $Q_0+1$ packets (including $P_1$) in slot $0.$ Observe
that $d_o,$ the separation between the probes at the output of the
queue, is affected by the arrivals and departures between the arrivals
of $P_1$ and $P_2$, i.e., by $A_{1,d_i}$ and $X_{0,d_i-1}.$ Thus, to
obtain the conditional distribution of the output separation, given
the input separation, we need to obtain the joint distribution of the
number of arrivals and departures between the two probe pairs.
Computation of this joint distribution is the key to the computation
of the distribution of the output separation.

From Eqn.~\ref{eq:one-queue-dispersion}, for a given $d_i$, say
$d_i=m,$ the distribution of the output separation $d_o$ can be
expressed as
\begin{eqnarray}
  \prob{d_o=s|d_i=m} & \ = \ & \sum_{l=1}^m \prob{X_{0,m-1}=l,
    A_{1,m}=s-m+l-1} 
  \label{eq:d_o-conditional}
\end{eqnarray}
for $s=1,2,\ldots.$.  Since we are assuming that we are probing a 
stationary queue, the first probe packet $P_1$ will see the queue in
steady state. We can then obtain the joint distribution of $X_{0,m-1}$
and $A_{1,m}$ as
\begin{eqnarray}
  && \prob{X_{0,m-1}=l, A_{1,m}=j } \  \ \nonumber \\
  && = \ \sum_{q=0}^{\infty}
  \prob{X_{0,m-1}=l, A_{1,m}=j | Q_0=q}  \prob{Q_0=q}  \nonumber\\ 
  && \ = \ \sum_{q=0}^{\infty}  \prob{X^q_{0,m-1}=l, A_{1,m}=j}
  \prob{Q_0=q} .
  \label{eq:X_m-A_a,m-joint-distbn}
\end{eqnarray}

We can see that $Q_0$ and $A_{1,m}$ are independent of each other, but
$X_{0,m-1}$ is dependent on both $Q_0$ and on the finite sequence
$A_1,\ldots, A_{m}.$ We make the reasonable assumption that the
distribution of $Q_0$ is known since it can be derived from the
distribution of $A_0$, albeit as a function of its parameters. Hence,
to obtain the distribution of $d_o,$ we first need to obtain the joint
distribution of $X_{0,m-1}$ and $A_{1,m}$ conditioned on $Q_0=q,$
i.e., the joint distribution of $X_{0,m-1}^q$ and $A_{1,m}.$ This is
derived in the following subsection.

\subsection{Joint Distribution of $A_{1,m}$ and $X_{0,m-1}^q$}
\label{sec:A_1,m-and-X_m-joint} 
In this section we first obtain a recursion on $m,$ for the joint
distribution of $A_{1,m}$ and $X^0_{0,m-1}$ and then, for arbitrary
$q,$ express the joint distribution of $A_{1,m}$ and $X^q_{0,m-1}$ in
terms of the joint distribution of $A_{1,m}$ and $X^0_{0,m-1}.$ We
thus begin with the case of $Q_0=0.$ From our notation, this means
that when the first probe packet is injected into the queue in slot
$0$ it is the only packet in the queue.  Since there is the probe
packet to be transmitted in slot $0,$ $X_{0,0}^0=1.$ Thus we have the
following base equations for the joint distribution.
\begin{eqnarray}
  \prob{X_{0,0}^0 = l, A_{1,1} = j} &=& 
  \begin{cases} 
    \prob{A_1 = j } & \mbox{for $l=1$},\\
    0 & \mbox{otherwise}.
  \end{cases} \nonumber \\
  \label{eq:base-equations}
\end{eqnarray}
Now, for arbitrary $m > 1,$ we can write the following one-step
recursion which also spells out the kind of recursion we seek. 

{\small
\begin{eqnarray}
  && \hspace{-0.35in} \prob{X_{0,m-1}^0=l , A_{1,m}=j } \  \ \nonumber \\ 
  && \hspace{-0.35in} = \ \sum_{i} \sum_{n} \prob{X_{0,m-1}^0=l,A_{1,m} =j |
    X^0_{0,m-2}=i, A_{1,m-1} = n} \nonumber \\  
  &&\hspace{0.4in} \times \, \prob{X_{0,m-2}^0=i, A_{1,m-1}=n}.
  \label{eq:basic-recursions}
\end{eqnarray}
}

\noindent
For ease of exposition, define
{\small
\begin{displaymath}
  \psi_{l,j}^{i,n}:= \prob{X_{0,m-1}^0=l,A_{1,m} =j | X^0_{0,m-2}=i,
    A_{1,m-1} = n} . 
\end{displaymath}
}
\normalsize

\noindent
We now consider three cases.
\begin{enumerate}
\item If $X^0_{0,m-2} = i < l-1$ or $X^0_{0,m-2} = i > l,$ then
  $\psi_{l,j}^{i,n} =0.$ This is because, for there to be $l$
  departures in slots $[0,m-1],$ the number of departures in slots
  $0,\ldots,(m-2)$, denoted by $i,$ should be either $l$ or $l-1.$

\item If $X^0_{0,m-2} = i = l-1$, then there has to be a departure in
  slot $m-1.$ Since $Q_0=0,$ this is possible if, and only if,
  $A_{1,m-1}=n \geq l-1$ because only then will the total number of
  arrivals in slots $[0,m-1],$ including $P_1,$ be greater than or
  equal to $l.$ Hence, for $n < l-1,$ $\psi_{l,j}^{i,n}=0$ and for
  $A_{1,m-1}=n \geq l-1,$ $X_{0,m-1}^0=l$ is a sure event and
  $A_{1,m}=j$ requires that $j-n$ packets have to arrive in slot $m$
  giving us $\psi_{l,j}^{i,n} = \prob{A_m=j-n}.$

\item For $X^0_{0,m-2} = i=l,$ there should be no departure in slot
  $m-1.$ Since $Q_0=0,$ $P_1$ has departed in slot $0.$ Hence the
  total number of departures in slots $[1,m-2],$ i.e., $l-1,$ should
  be the same as $A_{1,m-1}=n.$ Therefore, for $n \neq l-1,$
  $\psi_{l,j}^{i,n} = 0.$ Thus for $i=l,$ only $n = l-1$ will yield a
  non zero $\psi_{l,j}^{i,n}$ when the number of arrivals in slot $m$
  is $j-n=j-l+1,$ i.e., for $i=l$ and $n=l-1,$ we have
  $\psi_{l,j}^{i,n}=\prob{A_m=j-l+1}.$
\end{enumerate}
Combining the above three cases, we can rewrite the recursion in
Eqn.~\ref{eq:basic-recursions} for the joint distribution of
$X_{0,m-1}^0$ and $A_{1,m}$ as

{\small
\begin{eqnarray}
  && \hspace{-0.45in} \prob{X_{0,m-1}^0=l , A_{1,m}=j } \nonumber \\
  && \hspace{-0.45in} = \sum_{n=l-1}^{j}\prob{A_m=j-n} 
  \prob{X^0_{0,m-2}=l-1, A_{1,m-1} = n} 
  \nonumber \\ 
  && \hspace{-0.45in}  + \ \prob{A_m=j-l+1} \prob{X_{0,m-2}^0=l,
    A_{1,m-1} = l-1} . 
  \label{eq:recursions}
\end{eqnarray}
}

We obtain the joint distribution $X_{0,m-1}^q$ and $A_{1,m}$ by a
suitable transformation of the joint distribution of $X_{0,m-1}^0$ and
$A_{1,m}$ derived above. To obtain $\prob{X_{0,m-1}^q=l, A_{1,m}=j}$
we consider the following cases. Clearly, this probability is zero if
$l > m$ (the number of departures is greater than the number of slots)
and for $j < 0.$

\begin{enumerate}
\item The number of departures in slots $[0,m-1]$ is less than the
  number of slots, i.e., $l < m.$ We consider two sub~cases.
  \begin{enumerate}
  \item Clearly, $q > l-1$ is not possible because in that case, the
    number of departures including the probe packet $P_1$ would have
    been at least $l+1.$ Therefore, $\prob{X_{0,m-1}^q=l, A_{1,m}=j}
    =0$ for $q > l-1.$
  \item For $q \leq l-1,$ the event 
   \begin{displaymath}
     \{ ( A_1,\ldots,A_m) \ | \ X_{0,m-1}^q=l \}
   \end{displaymath}
    is the same as the event 
    \begin{displaymath}
      \{( A_1,\ldots,A_m) \ | \ X_{0,m-1}^0=l-q \}      
    \end{displaymath}
    and so 
    \begin{eqnarray*}
      && \hspace{-0.25in} \{(A_1,\ldots,A_m) \ | \ X_{0,m-1}^q=l,
      A_{1,m}=j\} \ = \ \\ 
      && \hspace{-0.1in} \{( A_1,\ldots,A_m) \ | \  X_{0,m-1}^0=l-q,
      A_{1,m}=j \}  
    \end{eqnarray*}
    Thus, for $q \leq l-1,$ we get
    \begin{eqnarray*}
      && \prob{X_{0,m-1}^q=l, A_{1,m}=j} = \\
      && \hspace{0.5in}\prob{X_{0,m-1}^0=l-q, A_{1,m}=j} 
    \end{eqnarray*} 
  \end{enumerate}

\item If $l=m$, then there is a departure in every slot in $[0,m-1]. $
  We once again consider two sub cases.
  \begin{enumerate}
  \item If $Q_0 =q < (l-1),$ then the event 
    \begin{displaymath}
      \{(A_1, \ldots, A_m) \ | \ X_{0,m-1}^q=m, A_{1,m}=j \}
    \end{displaymath}
    is the same as the union of the disjoint events 
    \begin{displaymath}
    \{(A_1, \ldots, A_m) \ | \ X_{0,m-1}^0 = t, A_{1,m}=j \}   
    \end{displaymath}
    for $t= m-q,\ldots, m.$ Hence, for this case, we can write
    \begin{eqnarray*}
      && \prob{X_{0,m-1}^q=l, A_{1,m}=j} \ = \ \nonumber \\
      && \hspace{0.4in} \sum_{t=m-q}^{m} \prob{X_{0,m-1}^0=t, A_{1,m}
        =j}. 
    \end{eqnarray*}

  \item If the queue starts with $Q_0 =q \geq (l-1)$ packets, then
    there will be a departure in every slot and $X_{0,m-1}^q=l$ is a
    sure event and so for this case $\prob{X_{0,m-1}^q=l, A_{1,m}=j} \
    = \ \prob{A_{1,m}=j}$.
  \end{enumerate} 
\end{enumerate}
Summarizing the above, we have the following transformation.
\begin{eqnarray}
  && \hspace*{-9mm}\prob{X_{0,m-1}^q=l, A_{1,m}=j}  \  \ \nonumber \\
  &&
  = \begin{cases}  
    \prob{X_{0,m-1}^0=l-q, A_{1,m}=j} & \\
    \hspace{0.9in} \mbox{for $l < m$ and $q \leq  l-1$} \\ 
    \sum_{t=m-q}^{m} \prob{X_{0,m-1}^0=t, A_{1,m} =j} & \\
    \hspace{0.9in} \mbox{for $l=m$ and $q < l-1$}  \\
    \prob{A_{1,m}=j} & \\
    \hspace{0.9in} \mbox{for $l=m$ and $q \geq l-1$} \\
    0 \hspace{0.9in} \mbox{otherwise.}
  \end{cases} 
  \label{eq:transformations}
\end{eqnarray}

Once the joint distribution of $X^q_{0,m-1}$ and $A_{1,m}$ is found
using the joint distribution of $X^0_{0,m-1}$ and $A_{1,m},$ the joint
distribution of $X_{0,m-1}$ and $A_{1,m},$ can be found from
Eqn.~\ref{eq:X_m-A_a,m-joint-distbn}.

The infinite summations in
Eqns.~\ref{eq:X_m-A_a,m-joint-distbn}~and~\ref{eq:d_o-unconditional}
will need to be truncated in practice. Let $N_1$ and $N_2$
respectively denote the truncation limits. Also, in
Eqn.~\ref{eq:d_o-conditional} we need to limit the range of the output
separation for which we will compute the probabilities. Let $N_3$ be
this limit. The algorithm for computing the distribution of the output
separation is summarized in Fig.~\ref{alg:boxed-algorithm}.  The
recursion in step 3b of Fig.~\ref{alg:boxed-algorithm} is
computationally the most intensive part of the algorithm. It can be
shown that this step requires
\begin{displaymath}
  \left(\frac{N_3(N_3+1)}{2} + \frac{5}{2} \right) \frac{N_2(N_2+1)}{2}
  - \frac{N_2(N_2+1)(2N_2 +1)}{12} 
\end{displaymath}
additions and multiplications. 

\small{
\begin{figure}
  \rule{6.3in}{2mm}
    {\small{
        \begin{enumerate}
        \item 
          \begin{tabbing}
            Input: \=Arrival distribution $\{p_k\}_{k\geq 0}$\\
            \>Input separation distribution $\{ \prob{d_i=k} \}_{k
              \geq 1}.$\\ 
          \end{tabbing}

        \item Obtain $\pi_q$ using the following equations.
          \begin{displaymath}
            \pi_q \ := \ \prob{Q_0=q} \ = \ 
            \begin{cases} 
              1 - \EXP{A_0}  \ = \ 1 - \lambda & \mbox{for $q=0$,} \\
              \frac{1}{p_0}\left( \pi_0(1-p_0) \right) & \mbox{for $q=1$,} \\
              \frac{1}{p_0}\left( \pi_{q-1}(1-p_1) - \pi_{q-2} p_2 -
                \cdots - \pi_1 p_{q-1} - \pi_0 p_{q-1}  \right) & \mbox{for $q >
                1$.} 
            \end{cases} 
          \end{displaymath}

        \item Obtain the joint distribution of $\{X^q_{0,m-1},
          A_{1.m}\}$ for each $q$ using the following equations.
          \begin{enumerate}
          \item Base Equations 
            \begin{eqnarray*}
              \prob{X_{0,0}^0 = l, A_{1,1} = j} &=& 
              \begin{cases} 
                \prob{A_1 = j } & \mbox{for $l=1$ and $j \ge 0$}\\
                0 & \mbox{otherwise}
              \end{cases}
            \end{eqnarray*}

          \item Recursions
            \begin{eqnarray*}
              \prob{X_{0,m-1}^0=l , A_{1,m}=j }
              &=&\sum_{n=l-1}^{j} \prob{A_m=j-n}
              \prob{X^0_{0,m-2}=l-1, A_{1,m-1} = n}  \nonumber \\   
              &&\hspace{-0.0in}  + \ \prob{A_m=j-l+1} \prob{X_{0,m-2}^0=l,
                A_{1,m-1} = l-1} . 
            \end{eqnarray*}
              
          \item Transformations
            \begin{displaymath}
              \prob{X_{0,m-1}^q=l, A_{1,m}=j}  = 
              \begin{cases}  
                \prob{X_{0,m-1}^0=l-q, A_{1,m}=j} & l < m \mbox{ and
                } q \leq  l-1 \\ 
                \sum_{t=m-q}^{m} \prob{X_{0,m-1}^0=t, A_{1,m} =j} &
                l=m \mbox{ and } q < l-1  \\
                \prob{A_{1,m}=j} & l=m \mbox{ and } q \geq l-1 \\
                0 & \mbox{otherwise}
              \end{cases}
            \end{displaymath}
          \end{enumerate}
        
        \item Obtain the joint distribution of $\{X_{0,m-1},
          A_{1,m}\}$ by using the following equations.
          \begin{displaymath}
            \prob{X_{0,m-1}=l, A_{1,m}=j } = \sum_{q=0}^{N_1}
            \prob{X^q_{0,m-1}=l, A_{1,m}=j} \pi_q
          \end{displaymath}

        \item Obtain the output separation distribution for a given
          input separation, $d_i,$ say $d_i=m.$
          \begin{displaymath}
            \prob{d_o=s|d_i=m} = \sum_{l=1}^m \prob{X_{0,m-1}=l,
              A_{1,m}=s-m+l-1} \hspace{0.5in} \mbox{for
              $s=1,2,\ldots,N_3$}.   
           \end{displaymath}

        \item Obtain the distribution of the output separation for a
          given input separation distribution.
          \begin{displaymath}
            \prob{d_o=s} =\sum_{j=1}^{N_2} \prob{d_o=s|d_i=j} \prob{d_i=j} . 
          \end{displaymath}
        \end{enumerate}
      }
    }
  \rule{6.3in}{2mm}
  
  \caption{Algorithm to compute the distribution of the output
    separation for a given distribution of the input separation and
    the arrival process.}
  \label{alg:boxed-algorithm}
\end{figure}
}

\section{Passage of a Probe Pair through a Network of Independent
  Queues}
\label{sec:probe-pair-network}
We now characterize the output separation between the probes as they
pass through multiple queues.  Since we are able to express the
distribution of the output separation of the probe pair in terms of
the distributions of the input separation and the arrival process
distribution, we can obtain the distribution of the output separation
at the output of a path of independent queues fairly easily.

\begin{figure}
  \begin{center}
    \     
    \includegraphics[width=4.5in]{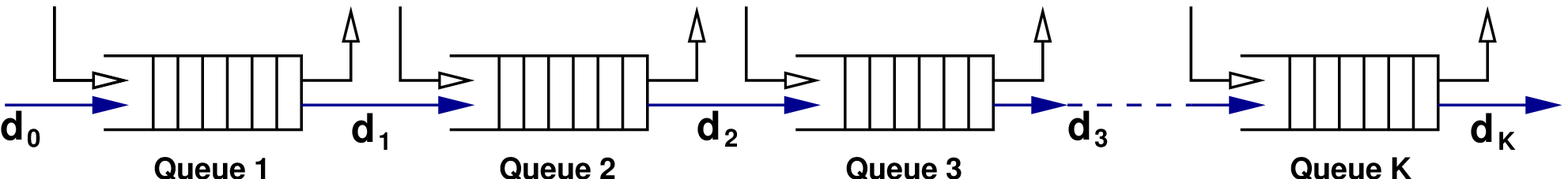}
  \end{center}
  \caption{$K$ independent queues and the path of the probe
    sequentially through the $K$ queues.}
    \label{fig:queue-series}
\end{figure}

We first consider the case when the probe pair passes through a path
consisting of $K$ independent queues as shown in
Figure~\ref{fig:queue-series}. The probe packets leaving queue~$k$ are
placed in queue $(k+1)$ after an arbitrary but fixed delay. Since the
delays are fixed, the separation between the probe pairs at the input
to queue~$(k+1)$ is the same as that at the output of queue~$k$.
(Generalization to variable delay is presented in
Section~\ref{sec:generalizations}.)  Let $d_0$ be the separation at
the input to queue~1 and $d_k$ the separation between the probe pairs
at the output of queue $k$ with $d_{\mathrm{out}}:=d_K.$

For $k=1, 2, \ldots, K,$ let $A_n^{(k)}$ denote the i.i.d.  arrival
sequence into queue $k,$ $p_i^{(k)},$ the distribution of $A_0^{(k)}$
and $\pi^{(k)}_i$ the stationary distribution of the queue occupancy
at the beginning of a slot.  Probe packets $P_1$ and $P_2$ are
injected into queue~1 in slots $0$ and $d_0$ respectively with a known
distribution of $d_0.$ The distribution of $d_k$ is obtained from the
recursive relation
\begin{equation}
  \prob{d_k=s} = \sum_{j=1}^\infty \prob{d_k=s |
    d_{k-1} =j} \prob{d_{k - 1} =j}. 
  \label{eq:d_k-distribution}
\end{equation}
The distribution of $d_K$ can be obtained by repeated use of the
algorithm in the previous section from the distribution of $d_0$.
Notice that, if $d_0=1,$ then $d_1$ has the same distribution as
$A^{(1)}_0.$

\begin{figure}
  \begin{center}
    \ 
    \includegraphics[width=3.5in]{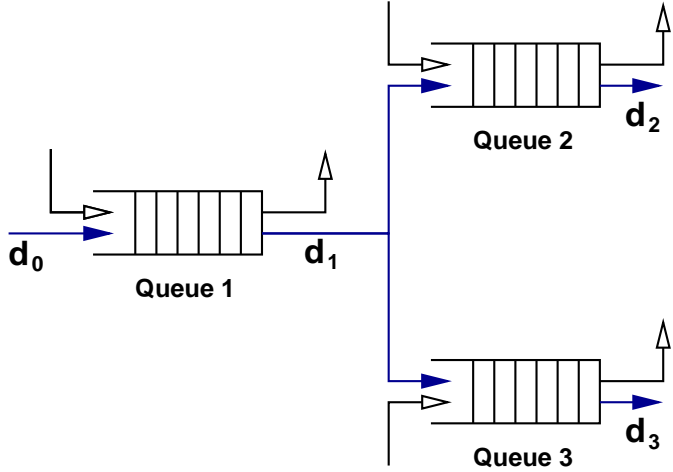}
  \end{center}
    \caption{A three-queue multicast tree.}
    \label{fig:queue-tree}
\end{figure}

The second case that is of interest is when the probe pairs traverse a
multicast tree of independent queues. Figure~\ref{fig:queue-tree}
shows an example of a three-node tree. Here copies of the probe packet
are made at the output of Queue~1 and the copies are simultaneously
placed in queues~2~and~3 after arbitrary but fixed delays on each of
the links. As with the case of a path, since the delays are fixed, the
separation between the probe pairs at the input to queues~2~and~3 is
the same as that at the output of queue~1. We are interested in the
joint distribution of the separation at the output of both
queues~2~and~3, $d_2$ and $d_3$ respectively. This joint distribution
is given by
\begin{eqnarray}
  && \hspace{-30pt} \prob{d_2=r, d_3=s} \nonumber  \\
  && \hspace{-30pt} \ = \ \sum_{t=1}^\infty \prob{d_2=r, d_3=s | d_1=t}
  \prob{d_1=t} \nonumber \\
  && \hspace{-40pt} \ = \ \sum_{t=1}^\infty  \prob{d_2=r|d_1=t}
  \prob{d_3=s|d_1=t} \prob{d_1=t} 
  \label{eq:multicast-joint-distbn}
\end{eqnarray}
The last equality follows because conditional on $d_1,$ $d_2$ and
$d_3$ are independent.  The individual conditional probabilities can
be obtained as before.

The above ideas can be extended to compute the joint distribution of
the separation at the leaves of an arbitrary tree rooted at the node
which introduces the probe pairs into the network.

\section{Estimating the Arrival Process Parameters}
\label{sec:estimating-A_n}

The algorithm developed for computing the distribution of the output
separation between the probes can, in principle, be used to estimate
the traffic parameters in the network provided the parameters are
identifiable from the distribution. However, often the parameters may
not be identifiable from the distribution, i.e., the same distribution
may be resulted from two different sets of traffic parameters. For
instanct, consider a series of two queues with independent Poisson
traffic with means $\lambda_1$ and $\lambda_2$. The output separation
has the same distribution for $(\lambda_1, \lambda_2) = (0,\lambda)$
and $(\lambda_1, \lambda_2) = (\lambda, 0)$. We performed some
simulation experiments to judge the viability of using this
distribution for traffic parameter estimation.  However, these
simulations use ideal settings like independent Poisson traffic
arrivals to different queues which are usually not strictly satisfied
by practical networks, and as a result, should be taken only as proof
of concept.  They do not guarantee that the method will work in
practical networks.

Consider a rooted tree of $K$ independent queues. Probe pairs are
introduced into this network at the root of the tree with unit
separation.  While the theory allows for an arbitrary but known
distribution of the input separation, we choose a fixed input
separation of one slot. This minimizes the computational requirements.

The separation between the probe pairs at the output of the leaves is
observed. We assume that there is no loss of probe pairs in the
network. We also assume that the link delays, i.e., the delay between
the output of a queue and the input to the next queue on a path is
arbitrary but fixed.

For $k=1,\ldots K,$ let $A^{(k)}_n$ be the arrival sequence of packets
into queue $k.$ For queue $k,$ $A^{(k)}_n$ is an i.i.d.~sequence and
further, the sequences for the different queues are independent.  Let
$\lambda_k$ be the vector of parameters of the distribution of
$A_0^{(k)}$ and let us define $\mathbf{\Lambda}:=[\lambda_1, \ldots,
\lambda_K].$ Our goal is to estimate $\mathbf{\Lambda}$ by observing
only the separation of the probes at the output of the leaves of the
paths.  Let $j_1, j_2, \ldots, j_{\kappa}$ denote these leaf nodes and
$d_{j_1},d_{j_2}, \ldots, d_{j_\kappa}$ the separation of the probe
pairs at the output of these queues.

We send a sequence of probe pairs and corresponding to every probe
pair we obtain an observation of the vector $\mathbf{d}:= [d_{j_1},
d_{j_2}, \ldots, d_{j_\kappa}].$ From these observations we can obtain
the empirical distribution of $\mathbf{d},$ denoted by
$\hat{\Psi}_{\mathbf{d}}$. We probe with a very low rate, i.e., the
interval between the probe pairs is made very large so that the
probing load is negligible.  This allows us to assume that the
stationary distribution of the occupancy of queue $k$ is governed only
by the distribution of $A_0^{(k)}$ and is a function of only
$\mathbf{\lambda}_k.$ Further, we assume that the first packet of
every probe pair, packet $P_1$, `sees' each of the queues in steady
state. This allows us to express the distribution of $\mathbf{d}$ as a
function of only $\mathbf{\Lambda}.$ We then use the results of the
previous sections to compute the analytical distribution
$\Psi_\mathbf{d} (\mathbf{\Lambda})$ for a given value of
$\mathbf{\Lambda}.$ Let $\mathcal{D}\left( \hat{\Psi}_{\mathbf{d}},
  \Psi_\mathbf{d} (\mathbf{\Lambda}), \right)$ denote a suitably
defined distance between $\hat{\Psi}_{\mathbf{d}}$ and
$\Psi_\mathbf{d} (\mathbf{\Lambda}).$ The estimate
$\hat{\mathbf{\Lambda}}$ of the parameter vector can be obtained as
\begin{displaymath}
  \hat{\mathbf{\Lambda}} = \arg \min_{\mathbf{\Lambda}} \ 
  \mathcal{D}\left( \hat{\Psi}_{\mathbf{d}}, \Psi_\mathbf{d}
    (\mathbf{\Lambda}) \right)  
\end{displaymath}
i.e., $\hat{\mathbf{\Lambda}}$ is obtained as the minimizer of the
distance between the empirical and the analytical distributions of the
separation vector.  

A well known measure of the distance between two distributions is the
Kullback-Leibler (KL) distance. The KL distance between the empirical
and the analytical distributions for a given $\mathbf{\Lambda},$
$\mathcal{D}_{\mathrm{KL}} \left( \Psi_\mathbf{d} (\mathbf{\Lambda}),
  \hat{\Psi}_{\mathbf{d}} \right),$ is defined as
\begin{eqnarray*}
  \mathcal{D}_{\mathrm{KL}} \left( \hat{\Psi}_{\mathbf{d}},
    \Psi_\mathbf{d} (\mathbf{\Lambda}) \right) := \sum_{\bf x}
  \hat{\Psi}_{\mathbf{d} = \mathbf{x}}  
  \log \frac{\hat{\Psi}_{\mathbf{d} = \mathbf{x}}}{\Psi_{\mathbf{d}=\mathbf{x}}
    \left(\mathbf{\Lambda}\right)}.
\end{eqnarray*}
It can be shown that the minimizer of $\mathcal{D}_{\mathrm{KL}}
\left( \hat{\Psi}_{\mathbf{d}}, \Psi_\mathbf{d} (\mathbf{\Lambda})
\right) $ is the maximum-likelihood estimate of $\mathbf{\Lambda}.$
Alternatively, we could use the squared Euclidean distance, as the
measure of the distance between the two distributions. The squared
Euclidean distance $\mathcal{D}_{\mathrm{E}} \left(
  \hat{\Psi}_{\mathbf{d}}, \Psi_\mathbf{d} (\mathbf{\Lambda}),
\right)$ between the empirical and the analytical distributions for a
given $\mathbf{\Lambda}$ is given by
\begin{eqnarray*}
  \mathcal{D}_{\mathrm{E}} \left( \hat{\Psi}_{\mathbf{d}},
    \Psi_\mathbf{d} (\mathbf{\Lambda}) \right)  := \sum_{\bf x} 
  \left( \hat{\Psi}_{\mathbf{d} = \mathbf{x}} -
    \Psi_{\mathbf{d}=\mathbf{x}}\left(\mathbf{\Lambda} \right) \right)^2 .
\end{eqnarray*}

A simplistic way to obtain $\hat{\mathbf{\Lambda}}$ is as follows. The
analytical distribution is computed and stored for all the values of
the parameter vector $\mathbf{\Lambda}$ over a suitably discretized
grid on the feasible range of $\mathbf{\Lambda}$. We obtain the
empirical distribution $\hat{\Psi}_{\mathbf{d}}$ from a sufficiently
large number of probes. The minimizer of $\mathcal{D}\left(
  \hat{\Psi}_{\mathbf{d}} , \Psi_\mathbf{d} (\mathbf{\Lambda})
\right)$ is then obtained by an exhaustive search over the grid.  To
find a reasonably accurate estimate, the exhaustive search will
require a very fine grid and consequently a large number of
distribution calculations. Rather than an exhaustive search, we can
also use an optimization method to reach the minimum of
$\mathcal{D}\left( \hat{\Psi}_{\mathbf{d}} , \Psi_\mathbf{d}
  (\mathbf{\Lambda}) \right).$

As we had mentioned earlier, for the case of one queue, if we fix
$d_0=1,$ then $d_1$ will have the same distribution as $A^{(0)}_1$ and
estimating the parameters of $A^{(0)}_1$ is straightforward. We will
not pursue that any more. The case of a two-queue path has interesting
numerical properties and we explore them in detail in the next
section. This will give us an insight into the difficulties of probing
paths that have multiple bottlenecks. We expand upon this in
Section~\ref{sec:generalizations}. We will also see that the
estimation quality when probing a multicast tree is significantly
better.

\section{Numerical Results}
\label{sec:numerical-results}
In this section we present some numerical results obtained as follows.
Different paths and multicast trees of independent discrete-time
queues are simulated. The probe packets are injected into the system
at very low probing rate. The passage of the probes through the
different queues are simulated and the output separation obtained for
each probe. We do not simulate lost probes although it is easy to see
that they do not affect the algorithm.
Although our results are
applicable for all distributions of the arrival process, we report our
results for the case when the packet arrivals form an i.i.d. Poisson
sequence.  Since Poisson is a one-parameter distribution, it is easier
to discuss the issues in the estimation of $\mathbf{\Lambda}.$

We first consider a two-queue path. Let $\lambda_1$ and $\lambda_2$ be
the arrival rates of packets to queues 1~and~2 respectively. Probe
pairs are generated according to a Poisson process. We obtain the
empirical distribution from $2 \times 10^8$ slots of simulation time.
Table~\ref{tbl:two-queue-exhaustive} shows the estimates that minimize
the squared Euclidean distance and KL distance between the empirical
distribution and the calculated distribution for different true values
of $\lambda_1$ and $\lambda_2$ and for two probing rates.  Notice that
both the distance measures give comparable precision in almost all
cases. Since the squared Euclidean distance is computationally simpler
than the KL distance, the rest of the numerical results are based on
the squared Euclidean distance.

{\footnotesize{
\begin{table*}
  \caption{Exhaustive search based estimates for the cascade of two
    queues}
  \label{tbl:two-queue-exhaustive}
  \begin{center}
  \begin{tabular}{||c||c||c|c|c||c|c|c||}
    \hline \hline
    True & inter &
    \multicolumn{6}{|c||}{Estimates in 3 different executions}\\
    \cline{3-8}
    values & probe & \multicolumn{3}{|c||}{Euclidean cost based}& 
    \multicolumn{3}{|c||}{KL distance based} \\
    \cline{3-8}
    $(\lambda_1, \lambda_2)$ & mean & $(\hat{\lambda}_1, \hat{\lambda}_2)$
    & $(\hat{\lambda}_1, \hat{\lambda}_2)$ & $(\hat{\lambda}_1, \hat{\lambda}_2)$
    & $(\hat{\lambda}_1, \hat{\lambda}_2)$
    & $(\hat{\lambda}_1, \hat{\lambda}_2)$ & $(\hat{\lambda}_1, \hat{\lambda}_2)$\\
    \hline \hline
    (0.9, 0.8) & 1000 & $(0.83, 0.85)$ & $(0.88, 0.81)$ & $(0.87, 0.82)$ & $(0.87, 0.82)$ & $(0.88, 0.81)$ & $(0.88, 0.81)$\\ \hline
    '' &  200  & $(0.91, 0.79)$ & $(0.91, 0.79)$ & $(0.87, 0.82)$ & $(0.87, 0.82)$ & $(0.88, 0.81)$ & $(0.88, 0.81)$\\\hline
    (0.5, 0.6) & 1000 & $(0.54, 0.57)$ & $(0.51, 0.59)$ & $(0.51, 0.59)$ & $(0.54, 0.57)$ & $(0.51, 0.59)$ & $(0.50, 0.60)$ \\ \hline
    '' &  200 & $(0.47, 0.62)$ & $(0.51, 0.59)$ &  $(0.49, 0.61)$ & $(0.51, 0.59)$ & $(0.50, 0.60)$ & $(0.49, 0.61)$ \\ \hline
    (0.7, 0.4) & 1000 & $(0.59, 0.50)$ & $(0.71, 0.39)$ &$(0.74, 0.36)$&
    $(0.68, 0.42)$ & $(0.71, 0.40)$ & $(0.68, 0.42)$ \\\hline
    '' &  200 & $(0.70, 0.40)$ & $(0.69, 0.41)$ & $(0.69, 0.41)$ & $(0.70, 0.40)$ & $(0.71, 0.39)$ & $(0.70, 0.40)$ \\ \hline
    (0.2, 0.3) & 1000 & $(0.22, 0.28)$ & $(0.13, 0.36)$ & $(0.35, 0.15)$ & $(0.22, 0.28)$ & $(0.13, 0.36)$ & $(0.21, 0.29)$ \\ \hline
    '' &  200 & $(0.20, 0.30)$ & $(0.19, 0.31)$ &$(0.21, 0.29)$& 
    $(0.20, 0.30)$ &$(0.20, 0.30)$ & $(0.20, 0.30)$ \\\hline
    \hline
  \end{tabular}
  \end{center}
\end{table*}
}}

The exhaustive search method has the drawback that it is primarily a
one time estimation and not an adaptive one. An alternative is to
reach the optimum value by a suitable optimization algorithm, for
example, a steepest descent algorithm. This may not however give the
best estimate if the distance function has a local minima. We
performed an extensive numerical investigation of the nature of the
cost function for the two-queue case. However, the cost function was seen to be non-convex. The level sets
of the Euclidian cost function as shown in Fig.~\ref{fig_contour} are
clearly not convex. This may present some problem in convergence of an
optimization technique. However, the limited numerical simulations
performed by us gave reasonably good performance under ideal setup.

\begin{figure}
  \begin{center}
    \ \includegraphics[height=2.70in]{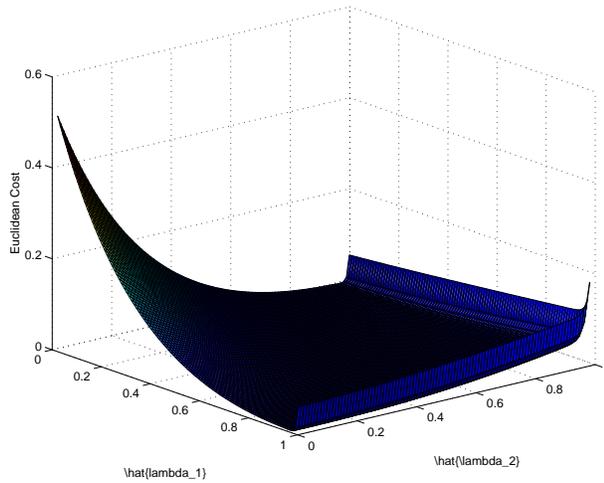}
  \end{center}
  \caption{A typical Euclidean cost as a function of $(\lambda_1,
    \lambda_2)$. This plot was obtained with $\lambda_1=0.3$ and
    $\lambda_2=0.8.$}
  \label{fig_cost}
\end{figure}
\begin{figure}
  \begin{center}
    \ \includegraphics[height=2.70in]{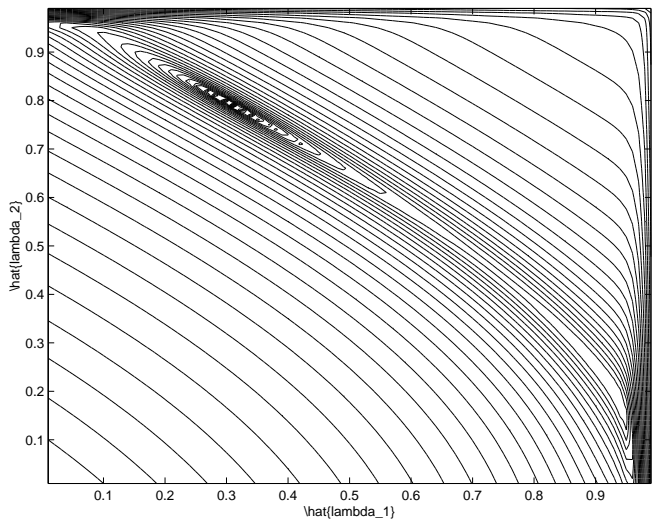}
  \end{center}
  \caption{Contours of the Euclidean cost function between the
    analytical and the empirical distributions for $\lambda_1=0.3$ and
    $\lambda_2=0.8.$ The contours were plotted by taking logarithm of
    the cost function so that more details can be seen near the valley
    (that almost touches zero).  }
  \label{fig_contour}
\end{figure}

Fig.~\ref{fig_cost} shows the plot of a typical Euclidean cost
function for an empirical distribution obtained from simulation with
$(\lambda_1, \lambda_2)=(0.3,0.8)$. The minima is at $(0.31,0.79)$.
However, notice that the function shows a distinct valley along a line
with an extremely low slope along it.  The existence of the valley is
more clear from the contour plot of the cost function shown in
Fig.~\ref{fig_contour}. The steepest descent method will result in a
very slow convergence due to zig-zagging along the valley.
 
Our main simulation involves an iterative stochastic gradient based
online estimation. We divide the probe sequence into equal sized
blocks and compute the `instantaneous' empirical distribution of
$\mathbf{d}$ for every block. In our simulations we have used a block
size of 500. Let $\hat{\Psi}^{\mathrm{inst}}_{\mathbf{d},n},$ be the
instantaneous empirical distribution obtained from the $n$-the block
of probes. The overall empirical distribution after the $n$-th
iteration, $\hat{\Psi}_{\mathbf{d},n},$ is updated using the
exponential update rule
\begin{eqnarray}
  \hat{\Psi}_{\mathbf{d},n} = 
  \begin{cases} 
    \hat{\Psi}^{\mathrm{inst}}_{\mathbf{d},1}& \text{if $n=1$} \\
    (1-a) \hat{\Psi}_{\mathbf{d},n-1} + a
    \hat{\Psi}^{\mathrm{inst}}_{\mathbf{d},n}  & \text{if $n>1$} .
  \end{cases}
  \label{update}
\end{eqnarray}
Here $a$ is a constant between $0$ and $1$. We used $a=0.05$ in all
our simulations. For each $n$, we then update our estimate of
$\mathbf{\Lambda}$ based on the current empirical distribution
$\hat{\Psi}_{\mathbf{d},n}$. We use the steepest descent algorithm for
one iteration for each $n$ and update the estimate of
$\mathbf{\Lambda}$ as
\begin{eqnarray}
  \hat{\mathbf{\Lambda}}_n = \hat{\mathbf{\Lambda}}_{n-1} -
  \alpha_n \frac{\nabla \mathcal{D}\left( \hat{\Psi}_{\mathbf{d},n},
      \Psi_\mathbf{d} (\mathbf{\Lambda}) \right) }{   
    \left| \nabla \mathcal{D}\left( \hat{\Psi}_{\mathbf{d},n} , \Psi_\mathbf{d}
        (\mathbf{\Lambda}) \right) \right| } 
  \nonumber
\end{eqnarray}
where $|\cdot|$ denotes the length of a vector and $\alpha_n$ is the
step size.  We used an adaptive step size so that whenever a
particular step increases the distance, the step size is decreased and
whenever the gradient is more than a threshold, the step size is
increased. The increase and decrease of the step size is not done if
it is respectively above or below some thresholds.  
If the arrival process is non~stationary, this scheme allows us to
track slowly varying $\mathbf{\Lambda}.$ Our simulations will study
the effectiveness of such tracking.

\begin{figure}
  \begin{center}
    \ \includegraphics[width=4.7in]{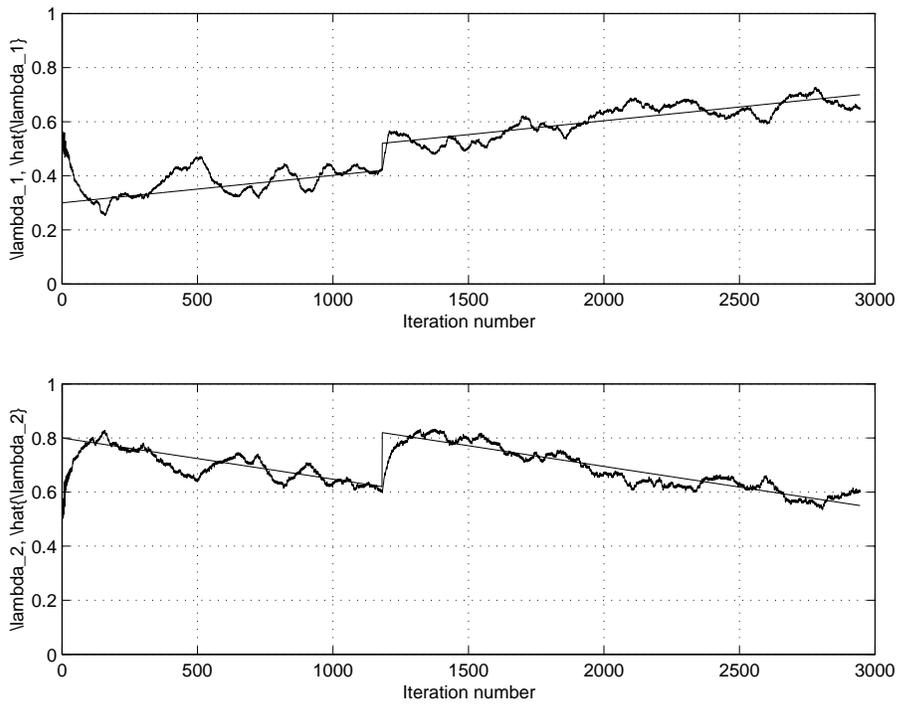}
  \end{center}
  \caption{The estimates of $\lambda_1$ and $\lambda_2$ and their true
    values for two-queue path. The straight lines represent the true
    values of the parameters.}
  \label{fig2q_a}
\end{figure}
\begin{figure}
  \begin{center}
    \ \includegraphics[height=2.70in]{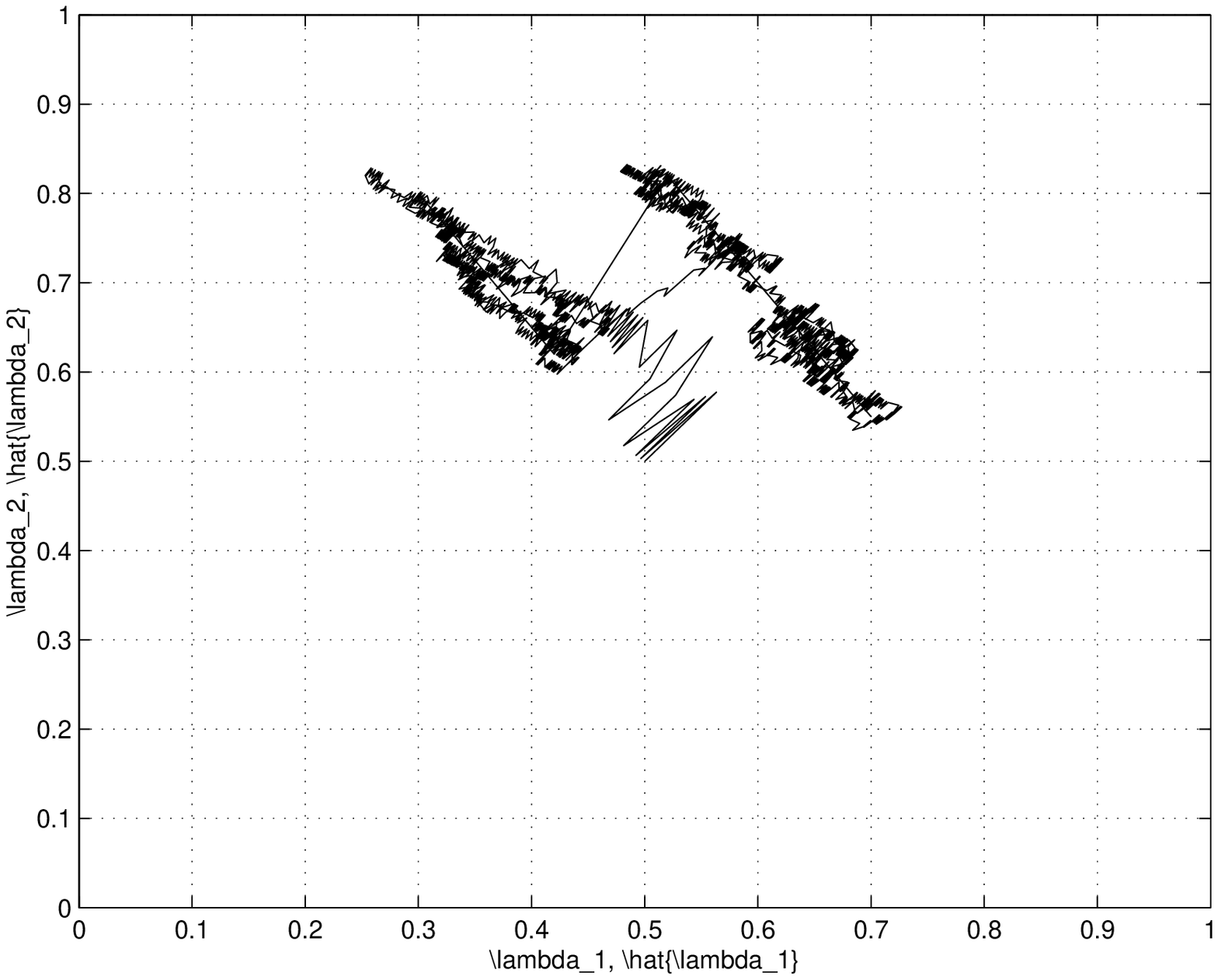}
  \end{center}
  \caption{The evolution of the estimates of $\lambda_1$ and
    $\lambda_2$ and their true values for the two-queue path. The
    straight lines show how the true value of $(\lambda_1, \lambda_2)$
    changes.  }
  \label{fig2q_b}
\end{figure}
We first consider the two-queue path. $\lambda_1$ and $\lambda_2$ are
varied linearly with time and a step change is also simultaneously
introduced. Fig.~\ref{fig2q_a} plots the true values of $\lambda_1$
and $\lambda_2$ and the estimates as functions of time.  For the
estimation, the initial estimates are taken as $\hat{\lambda}_1 =
\hat{\lambda}_2=0.5.$ From these plots we see that the steepest
descent algorithm converges and tracks the true values reasonably well
even when there is step change. It is instructive to study the
evolution of the estimate. Fig.~\ref{fig2q_b} shows the evolution of
the estimate pair $(\hat{\lambda}_1, \hat{\lambda}_2).$ A close look
at the figure reveals that initially as well as when the parameters
change suddenly, the estimates go to the nearest point on the new
valley and then zig-zags along the valley towards the true values.

\begin{figure}
  \begin{center}
    \
    \includegraphics[width=4.5in]{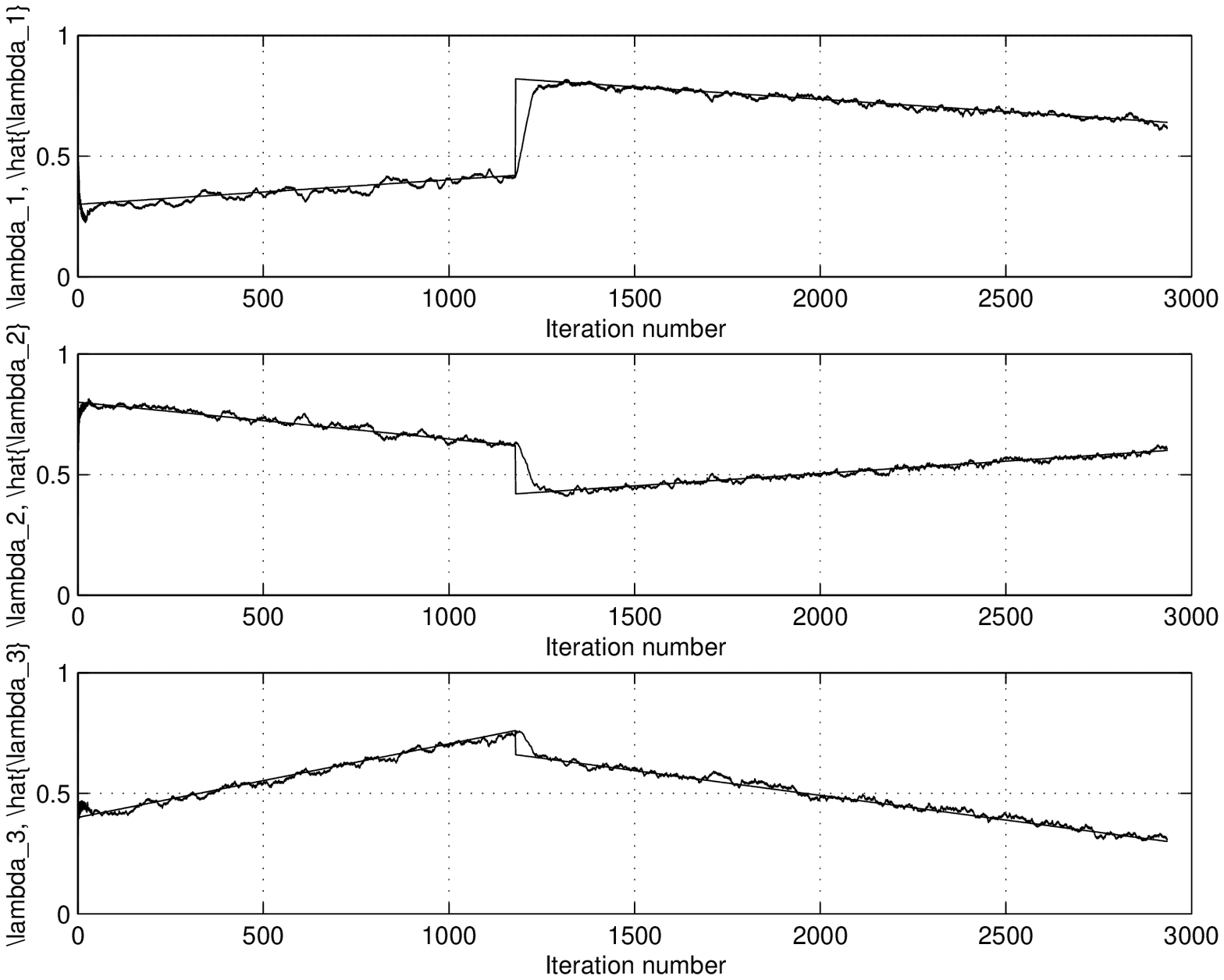}
  \end{center}
  \caption{The estimates of $\lambda_1$, $\lambda_2$ and $\lambda_3$
    and their true values for the multicast network}
  \label{fig_mult}
\end{figure}

We next consider the three-queue multicast tree of
Fig.~\ref{fig:queue-tree}. As with the two-queue case, the true values
of $\lambda_i$ are varied linearly and also a step change is
introduced. In Fig.~\ref{fig_mult} we plot $\hat{\lambda}_1$,
$\hat{\lambda}_2$ and $\hat{\lambda}_3$ as functions of time along
with the true values. Observe that these estimates are far superior to
those for the two-queue path. This is because here we have the
empirical joint distribution of the output separations at two leaf
nodes of the network and this gives much more information about the
arrivals to the queues.

The slow convergence of the adaptive estimation due to the valley in
the Euclidean cost function that we saw for the two-queue path is
expected to be even more pronounced for a three-queue path. In fact,
in our simulations, the estimates did not converge to the true values
even after $2 \times 10^8$ slots of simulation time.  We conjecture
that there will be a valley along a two dimensional manifold.

\begin{figure}
  \begin{center}
    \ 
    \includegraphics[width=3.2in]{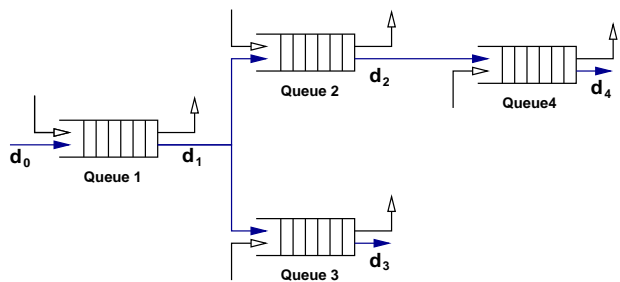}
  \end{center}
  \caption{The four-queue multicast tree used in some of the
    simulation results. }
  \label{fig:four-queue-tree}
\end{figure}

\begin{figure}
  \begin{center}
    \
    \includegraphics[width=4.5in]{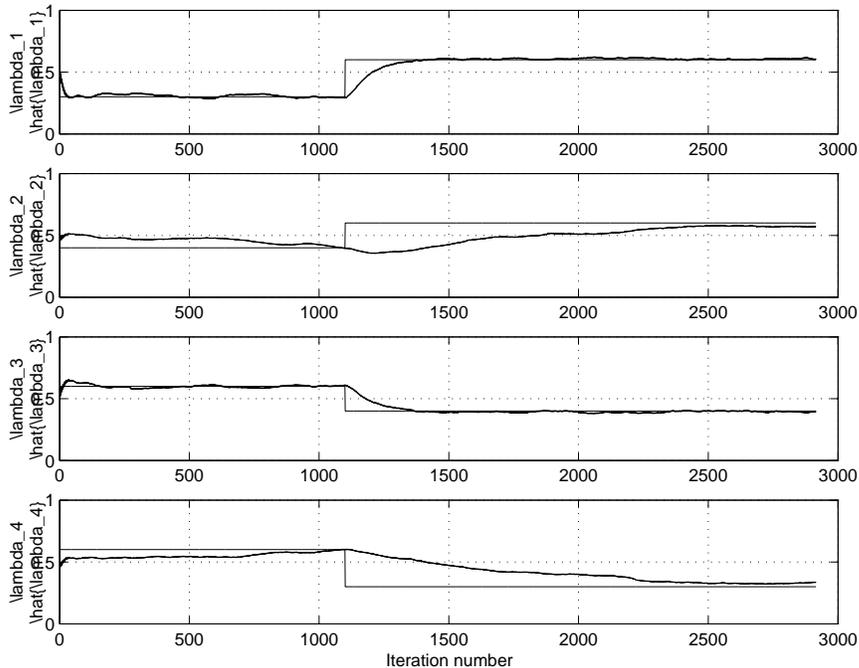}
  \end{center}
  \caption{The estimates of $\lambda_1$, $\lambda_2$, $\lambda_3$ and
    $\lambda_4$ and their true values for the network of
    Fig.~\ref{fig:four-queue-tree}.}
  \label{fig_4qnet}
\end{figure}

We have seen that in a multicast network, having multiple outputs
through different paths helps improve the quality of the estimation.
However two or more queues in series makes for slow convergence of the
estimates. To investigate the interplay of these two we carried out
the estimation for a four-queue multicast tree shown in
Fig.~\ref{fig:four-queue-tree}.  The plots of the estimates and the
true values of the arrival means are shown in Fig.  \ref{fig_4qnet}.
As expected, we observed faster convergence for this network. However,
the queues~2~and~4 are in series in this network. The estimates of
$\lambda_2$ and $\lambda_4$ can be seen to be worse than the estimates
for the two-queue path.  This is possibly because of the error in the
estimate of $\lambda_1$ (this misleads the estimation of $\lambda_2$
and $\lambda_4$ with an incorrect distribution of the probe separation
at the input of queue~2) and also because the probe pairs entering
queue~2 are not separated by a fixed one slot. Hence, the convergence
of the estimates of $\lambda_2$ and $\lambda_4$ is not as fast as that
of $\lambda_1$ and $\lambda_3$.

In the adaptive estimation, we need to consider two performance
criteria. If the arrivals are stationary, we need to minimize the
error in the estimate. However, if the arrivals are non-stationary, we
would like the estimates to track the true values closely. Achieving
these objectives is governed by the choices of the various parameters
of the estimation algorithm.  In the rest of this section we will
discuss the role of these parameters. The performance of the estimation
was seen to be quite sensitive to the selection of the parameters,
and quite a bit of tuning was necessary to get the performance
shown in the figures.

\noindent
\emph{Choice of $a,$ block length and $\alpha_n$:} The parameter $a$
(in Eqn.~\ref{update}) and the number of probe pairs in a block
dictate how fast the empirical distribution tracks the true
distribution of the output separations.  However, there is a trade-off
between tracking the true distribution for non-stationary arrivals and
the variance of the empirical distribution under stationary arrivals.
Increasing $a$ or decreasing the block size enables the empirical
distribution to track the actual distribution faster but increases the
variance of the empirical distribution under stationary arrivals. This
reflects in $\hat{\mathbf{\Lambda}}_n$ tracking the true value faster
but in the components of $\hat{\mathbf{\Lambda}}_n$ having larger
variance under stationary arrivals.

The step size $\alpha_n$ also plays a role in the convergence rate and
the tracking ability of the estimation. A larger $\alpha_n$ enables
faster tracking but results in larger variance of the estimates under
stationary arrivals.

\noindent
\emph{How well to track the minima of $\mathcal{D} \left(
    \hat{\Psi}_{\mathbf{d},n}, \Psi_{\mathbf{d}}\left(
      \mathbf{\Lambda} \right) \right)$ :} In our simulations, we have
used one iteration of steepest descent for each $n$.  Alternatively we
could use an exhaustive search to obtain the minima of $\mathcal{D}
\left( \hat{\Psi}_{\mathbf{d},n}, \Psi_{\mathbf{d}}\left(
    \mathbf{\Lambda} \right) \right).$ As a second alternative, we may
execute many iterations of steepest descent for each $n$. This may
involve a fixed number of iterations or termination based on a
specified condition.  A larger number of iterations for each $n$ will
enable $\hat{\mathbf{\Lambda}}_n$ to be nearer the minima of the
current cost function $\mathcal{D}\left( \hat{\Psi}_{\mathbf{d},n},
  \Psi_\mathbf{d} (\mathbf{\Lambda}) \right).$ Hence, a larger number
of iterations for each $n$ will result in the estimate
$\hat{\mathbf{\Lambda}}_n$ tracking the minima of $\mathcal{D}\left(
  \hat{\Psi}_{\mathbf{d},n}, \Psi_\mathbf{d} (\mathbf{\Lambda})
\right)$ more closely.  Although this results in better tracking, it
makes the estimate more tuned to the current empirical distribution
and as a result more sensitive to the variation of the current
empirical distribution under stationary arrivals.

\noindent
\emph{Alternative Optimization Methods:} The valley in the cost
function makes the convergence of steepest descent algorithm very slow
due to the typical zig-zag path followed by this algorithm.  Other
optimization algorithms like conjugate gradient or Newton's algorithm
may also be tried to overcome the problem.

\noindent
\emph{Method of Moments Based Estimation:} A common method in
parameter estimation is the method of moments.  A similar approach can
be considered here. We can compute a few moments of the calculated and
the empirical distribution and try to match them by varying the
parameters. In the absence of closed form expressions for the moments,
we may again need to estimate by minimizing a suitable cost function
between the calculated moments and the moments from the empirical
distributions. However, since we do not have a way of calculating the
moments directly without calculating the distributions first, this
method presents little attraction, especially since we can expect to
lose some information in working only with some moments instead of the
distributions themselves.  Indeed, the results obtained by using the
squared Euclidean distance between the vectors of the first ten
moments of the calculated and the empirical distributions gave no
promising results.  It is also to be noted that computation of higher
order moments is computationally intensive and needs numerical caution
in dealing with large numbers.

\noindent
\emph{Probing rate:} Finally, we remark on the probing rate. A higher
probing rate gives us more samples and hence a quicker convergence for
stationary arrivals and a better ability to track fast changing
arrival processes. However, the probing process makes the aggregate
arrival process, including the probe packets, non Poisson and disturbs
the estimate because the calculated distribution will not be correct.
Further, the samples may not be independent if the probes are closely
spaced. To minimize the effect, the mean probe arrivals should be
small. In our simulations, the times at which probes $P_1$ are
inserted were generated according to a Poisson process of rate
$0.005.$

\section{An Extension and a Generalization}
\label{sec:generalizations}
In this section we consider an extension and a generalization to the
algorithms presented in
Sections~\ref{sec:output-dispersion}~and~\ref{sec:probe-pair-network}.
First, we relax the assumption that the probe packet has the least
priority within its batch. We will see that this essentially involves
extension of the joint distribution of $A_{1,m}$ and $X_{0,m-1}^q$ of
Section~\ref{sec:A_1,m-and-X_m-joint} to include packets that arrive
in the same slot as the probes but may be queued after the probe
packet.  We then present a generalization in which we allow the
transfer delays in the network to be random rather than fixed but
arbitrary.

\subsection{Generalization to Equal Priority for the probes}
\label{sec:equal-priority}
In our discussions earlier, we assumed that the probe packets have the
lowest priority among the packets that arrive in the slot. We now
consider the case when all the packets that arrive in a slot have the
same priority and the position of the probe packets within the batch
may have a distribution.

\begin{figure}
  \begin{center}
    \ \includegraphics[width=4.5in]{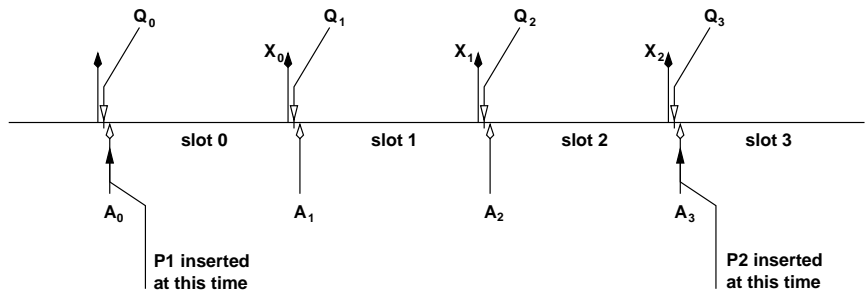}
  \end{center}
  \caption{The convention for $A_n,$ $X_n$ and $Q_n$ used in
    Section~\ref{sec:equal-priority}. }
  \label{fig:general-convention}
\end{figure}

For convenience, we redefine the notation and let $Q_n$ be the number
of packets in the queue at the beginning of slot $n$ but \emph{before}
the arrival of the packets in slot $n$. This new notation is shown in
Figure~\ref{fig:general-convention}. Under this notation the queue
evolution described by Eqn.~\ref{eq:queue-evolution} changes to
\begin{eqnarray}
  Q_{n+1} &=& Q_n + A_{n} - X_{n} \ ,\nonumber \\
  Q_{n+m} &=& Q_n + A_{n,n+m-1} - X_{n,n+m-1} \ .
  \label{eq:general-queue-evolution}  
\end{eqnarray}

Note that according to this definition the stationary distribution of
$Q_n$ would be different from that given in
Eqn.~\ref{eq:pi_q-recursion}.  The probability generating function for
the stationary distribution would be
\begin{displaymath}
  \mathcal{Q}(z) = \frac{(1-\lambda)(z-1)}{z-\mathcal{P}(z)} 
\end{displaymath}
and the recursion for the probabilities will be
\begin{displaymath}
  \pi_q \ = 
  \begin{cases}
    \frac{1-\lambda}{p_0} & \mbox{for $q=0$} \\
    \frac{\pi_0}{p_0}\left(1-p_0 - p_1\right) & \mbox{for $q=1$} \\
    \frac{1}{p_0}\left( \pi_{q-1}(1-p_1) - \pi_{q-2} p_2 \right. & \\
      \hspace{0.2in}\left. - \cdots - \pi_0 p_{q} \right) & \mbox{for
        $q > 1$} . 
  \end{cases}
\end{displaymath}

For a probe packet arriving into the queue in slot $k,$ define
$A^\prime_k$ (resp.~$\tilde{A}_k$) to be the number of packets from
those that arrived in slot $k$ that are queued before (resp.~after)
the probe packet. Thus $A^\prime_k + \tilde{A}_k = A_k.$ Also, for
probe packets arriving in slots $i$ and $j,$ define,
\begin{eqnarray*}
  A_{i,j}^\prime &:=& A_i + A_{i+1} + \cdots + A_{j-1} + A^\prime_j \ \,\\
  \tilde{A}_{i,j} &:=& \tilde{A}_i + A_{i+1} + \cdots + A_{j-1} +
  A_j\ , \\
  \bar{A}_{i,j} &:=& \tilde{A}_i + A_{i+1} + \cdots + A_{j-1} +
  A^\prime_j \ . 
\end{eqnarray*}

As before, we assume that the probe packets $P_1$ and $P_2$ enter the
queue in slots $0$ and $d_i$ respectively. They depart in slots $D_1$
and $D_2$. Along the lines of Eqn.~\ref{eq:one-queue-dispersion} we
obtain the output separation of the probe packets as follows.  
\begin{eqnarray}
  D_1 &=& Q_0 + A^\prime_0 \ , \nonumber \\
  D_2 &=& d_i + Q_{d_i} + A^\prime_{d_i} \nonumber \\
      &=& d_i + Q_0 + 1 + A_{0,d_i-1} -  X_{0,d_i-1} + A_{d_i}^\prime
      \ ,  \nonumber \\  
  d_{\mathrm{out}} &:=& D_2 - D_1 \nonumber \\
  &=& d_i + 1 + \bar{A}_{0,d_i} - X_{0,d_i-1}  \ . 
  \label{eq:general-dispersion}
\end{eqnarray}

For any $d_i=m+1 > 1,$ to obtain the distribution of the output
separation, we thus need to know the joint distribution of $X_{0,m}$
and $\bar{A}_{0,m+1}.$

The distribution of $A_m^\prime$ and $\tilde{A}_m$ is derived from the
distribution of $A_0$ as
\begin{displaymath}
  \prob{A^\prime_m=k} = \prob{\tilde{A}_m=k}  = \sum_{j=k}^\infty
  \frac{1}{j+1} \prob{A_0=j}.  
\end{displaymath}
Following the procedure in obtaining the recursion in
Section~\ref{sec:probe-pair-network} we begin by defining
$X_{0,m}^{q,a}$ to be the number of departures in slots $0,\ldots,m$
when $Q_0=q$ and $A_0=a.$ 

As before, we begin by assuming $Q_0=A_0=0$ and compute the joint
distribution of $X_{0,m}^{0,0}$ and $A_{1,m}^\prime.$ The base
equation will be
\begin{eqnarray}
  \prob{X_{0,0}^{0,0} = l, A_{1,0} = j} &=& 
  \begin{cases} 
    1 & \mbox{for $l=1$ and $j=0$,}\\
    0 & \mbox{otherwise.}
  \end{cases} \nonumber \\
  \label{eq:base-equations-prime}
\end{eqnarray}
Here $A_{1,0}$ is an empty summation and we define it to be 0.  We
first obtain the recursion for the joint distribution of
$X^{0,0}_{0,m}$ and $A_{1,m}$ and then the joint distribution of
$X^{0,0}_{0,m}$ and $A^\prime_{1,m+1}.$ Following the arguments of
Section~\ref{sec:output-dispersion}, we obtain the following
recursions.
\begin{eqnarray}
  && \hspace{-0.2in}\prob{X_{0,m}^{0,0}=l , A_{1,m}=j }  \ \nonumber \\
  && \hspace*{-.1in} =
  \begin{cases}
    \prob{X_{0,m-1}^{0,0}=l , A_{1,m-1}=l-1 }  \\
    \hspace{0.7in} \times \prob{A_m=0} \ \delta(j-l+1)  \\
    \hspace{0.1in} \ + \ \sum_{n=l-2}^j \prob{X_{0,m-1}^{0,0}=l-1,
      A_{1,m-1}=n} \\
    \hspace{0.7in} \times \prob{A_m=j-n} ,\\
    \hspace{0.9in} \mbox{for $l \leq m+1,$ and $j > l-2$,} \\
    0  \hspace{0.9in} \mbox{otherwise}
  \end{cases} \\
  &&\hspace{-0.2in}\prob{X_{0,m}^{0,0}=l , A^\prime_{1,m+1}=j }  \ = \
  \nonumber \\ 
  && \sum_{k=0}^j \prob{X_{0,m}^{0,0}=l , A_{1,m}=j-k }
  \prob{A^\prime_{m+1} =k} , 
  \label{eq:recursions-prime}
\end{eqnarray}
where $\delta(\cdot)$ is the Kronecker delta function. If $Q_0 =q$ and
$A_0=a,$ then, following the arguments in the development of
Eqn.~\ref{eq:transformations}, the joint distribution of
$X^{q,a}_{0,m}$ and $A^\prime_{1,m+1}$ would be obtained in exactly
the same manner as Eqn.~\ref{eq:transformations}, except that $q$
would be replaced by $q+a$.  Thus we will have
\begin{eqnarray}
  && \hspace*{-0.3in}\prob{X_{0,m}^{q,a}=l, A^\prime_{1,m+1}=j}   \nonumber \\
  && = 
  \begin{cases}  
    \prob{X_{0,m}^{0,0}=l-q-a, A^\prime_{1,m+1}=j} \\
    \hspace{0.5in} \mbox{for $l < m+1$ and $q+a \leq l-1,$} \\
    \sum_{t=m+1-q-a}^{m+1} \prob{X_{0,m}^{0,0}=t, A^\prime_{1,m+1} =j}
    \\
    \hspace{0.5in} \mbox{for $l=m+1$ and $q+a < l-1,$}  \\
    \prob{A^\prime_{1,m}=j} \\
    \hspace{0.5in} \mbox{for $l=m+1$ and $q+a \geq l-1,$} \\
    0  \hspace{0.4in} \mbox{otherwise.}
  \end{cases}
  \label{eq:transformations-prime-1}
\end{eqnarray}
Of the $a$ packets that arrive in slot $0$, if $\tilde{a}$ are queued
after $P_1,$ then we need that many fewer arrivals in $1, \ldots, m.$
Hence, we have
\begin{eqnarray}
  && \hspace{-0.4in} \prob{X_{0,m}^{q,a}=l, \bar{A}_{0,m+1}=j} \  \
  \nonumber \\ 
  && = \ \sum_{\tilde{a}=0}^a \frac{1}{a+1} \prob{X_{0,m}^{q,a}=l,
    A^\prime_{1,m+1}=j-\tilde{a} }. 
  \label{eq:transformations-prime-2}  
\end{eqnarray}
Finally, we can uncondition on $q$ and $a$ as
\begin{eqnarray}
  && \prob{X_{0,m}=l, \bar{A}_{0,m+1}=j} \ = \sum_{q=0}^{\infty}
  \sum_{a=0}^\infty  \prob{A_0=a} \nonumber \\
  && \hspace{0.4in} \prob{Q_0=q} \prob{X_{0,m}^{q,a}=l,
    \bar{A}_{0,m+1}=j} . 
  \label{eq:uncondition-on-a}  
\end{eqnarray}

The output separation of the probes can now be obtained following the
same method as in Section~\ref{sec:output-dispersion}.

\subsection{Random Delays on the Links}
\label{sec:random-link-delays}
In deriving the distribution of the output separation, in
Eqns.~\ref{eq:d_k-distribution} and \ref{eq:multicast-joint-distbn},
we assumed that the transfer delays from the output of a queue to the
input of the next queue on the path is arbitrary but fixed. We can
relax this requirement as follows.

Consider the series of queues of Figure~\ref{fig:queue-series} first.
In going from the output of queue $k$ to the input of queue $k+1$, let
$P_1$ be delayed by $t_1$ slots and $P_2$ by $t_2$ slots. Let $d_k$ be
the probe separation at the output of queue $k.$ The probe separation
at the input to queue $k+1$ is $d^\prime_k := d_k + (t_2 - t_1).$ It
is reasonable to assume that the order of the probes is maintained in
the transfer from $k$ to $k+1$. Hence $t_1$ and $t_2$ cannot be
independent. If the conditional distribution of $d_k^\prime$ given
$d_k$ is known then Eqn.~\ref{eq:d_k-distribution} can be generalized
as 
\begin{eqnarray*}
  && \prob{d^\prime_{k-1} =j} \ = \ \sum_{i=1}^\infty
  \prob{d_{k-1}^\prime=j 
    | d_{k-1} =i} \\
  && \hspace{-0.1in}
   \prob{d_k=s} \ = \ \sum_{j=1}^\infty \prob{d_k=s | d^\prime_{k-1}
     =j} \prob{d_{k-1}^\prime=j}  
\end{eqnarray*}

\begin{figure}
  \begin{center}
    \
    \includegraphics[height=1.5in]{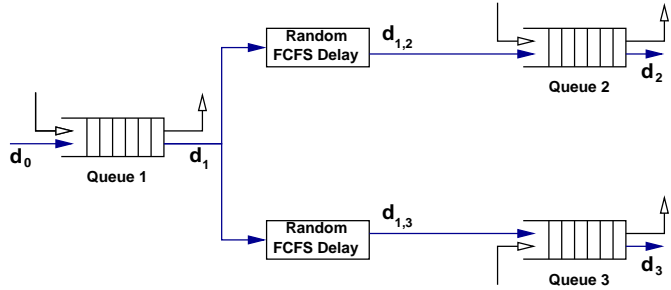}
  \end{center}
  \caption{The multicast tree with random transfer delays. }
  \label{fig:tree-random-transfer-delay}
\end{figure}
Along similar lines, we can generalize for random transfer delays on
the multicast tree. The schematic is shown in
Fig.~\ref{fig:tree-random-transfer-delay}.
Eqn.~\ref{eq:multicast-joint-distbn} can be generalized as
\begin{eqnarray*}
  && \hspace{-0.2in} 
  \prob{d_2=r, d_3=s} \nonumber \\
  &&  \hspace{-0.2in} = \ \sum_{n=1}^\infty \sum_{m=1}^\infty 
  \sum_{l=1}^\infty \prob{d_2=r|d_{1,2}=l} 
  \prob{d_3=s|d_{1,2}=m} \nonumber \\
  && \hspace{0.5in} \times \ \prob{d_{1,2}=l; d_{1,3}=m| d_1=n}
  \prob{d_1=n}  
\end{eqnarray*}

There are many uses for this generalization. If the transmission rates
on the links are different, then the separation is compressed or
expanded depending on whether the downstream link has a higher or
lower data rate respectively. The conditional distribution of the
separation can be obtained by knowing the two transmission rates.

A second use is in tomography of paths with many links with one or two
of them being `bottleneck links.' Consider a $K$-queue path.  Let
links $i$ and $j,$ $1 \leq i < j \leq K,$ be the bottleneck links.
The delays from the source to the input of link $i$, from the output
of $i$ to input of $j$ and from the output of $j$ to the destination
can have simple models. For example, let $K=3,$ $i=1$ and $j=3$ with
link $2$ being lightly loaded. We can assume that over link~2, each of
the probes $P_1$ and $P_2$ will see at most one cross traffic packet
independently. Let $\alpha$ be the probability of this event. Then we
can see that, if the input separation is greater than 1, the
separation increases by 1 if only $P_2$ is sees a cross traffic
packet, while it shrinks by 1 if only $P_1$ sees such a packet. If the
input separation is~1, then the output separation will also be~1 if
neither $P_2$ does not see a cross traffic packet. We can then use
\begin{eqnarray*}
  && \hspace{-0.25in}\prob{d_2=n|d_1=m} \  \\
  && \hspace{-0.25in}=
  \begin{cases}
    \alpha & \mbox{if $m=1$ and $n =2 $} \\ 
    1-\alpha & \mbox{if $m=n=1$} \\ 
    \alpha(1-\alpha) & \mbox{if $n = m+1$ or $n=m-1$ \& $m>1$} \\
    1-2\alpha(1-\alpha) & \mbox{if $n=m$ and $m > 1$} \\ 
    0 & \mbox{otherwise}.
  \end{cases}
\end{eqnarray*}
Here link~2 is playing the role of the random delay element. 

\section{Discussion and Conclusion}
\label{sec:discussion}
We conclude with a discussion on the connection with some earlier
results on the analysis of packet-dispersion based probing and also on
network delay tomography.

We first explore the connection of this work with that in
\cite{Liu05,Liu07}. A theoretical analysis of packet pair probing of a
single bottleneck path is carried out in \cite{Liu05}. A continuous
time model is used and the key result is Theorem~2. This theorem
states that for a probe pair inserted into the queue at time $a_1$
with input separation $\delta$ and probe packet length $s,$ the output
separation $\delta^\prime$ is given by
\begin{equation}
  \delta^\prime  =  \frac{Y_\delta(a_1)}{C} + \frac{s}{C} +
  \max \left( \frac{B_\delta(a_1)  -s }{C}, 0 \right)
  \label{eq:loguinov-thm2}
\end{equation}
where  
\begin{itemize}
\item $C$ is the capacity of the bottleneck link,
\item $Y_\delta(t)$ is the work (sum of the service times of the
  packets) that enters the queue in $(t, t+\delta)$ and 
\item $B_\delta(t)$ is the unused capacity of the server in the
  interval $(t, t+\delta).$ 
\end{itemize}
Thus the output separation is a sample of a linear combination of
three interdependent random processes. This theorem essentially
implies that obtaining the joint law of these processes is the key to
the use of packet pair probing.

We can interpret Eqn.~\ref{eq:loguinov-thm2} for the discrete time
queues that we consider as follows. Note that $s=C=1.$ The first term
is the number of packets entering the queue between the enqueuing of
the two probe packets and the third term is the number of idle slots
between the enqueuing of the two probe packets. Thus we can rewrite
Eqn.~\ref{eq:loguinov-thm2} as follows.
\begin{eqnarray*}
  d_{\mathrm{out}} &=& A_{1,d_i} + 1 + \max \left\{ d_i - X_{0, d_i-1}, 0
\right\}\\   
&=& A_{1,d_i} + 1 + d_i - X_{0, d_i-1} .
\end{eqnarray*}
The last equality is true because the number of departures in $d_i$
slots can never be greater than $d_i.$ This is nothing but
Eqn.~\ref{eq:one-queue-dispersion}!

Eqn.~\ref{eq:loguinov-thm2} essentially says that the output
separation of a packet pair is a function of the work entering the
queue between the probe arrivals and the amount of work done in this
interval. These two quantities are dependent and one needs their joint
distribution for finding the distribution of $d_{\mathrm{out}}.$ The
key difficulty in developing the analytical model for packet pair
probing is to obtain this joint distribution, and we believe, that is
a key contribution of this paper.

It is interesting to contrast our problem setting with delay
tomography of, for example, \cite{Coates01,Tsang01}, since both
estimate how the individual links will affect the traffic flowing,
albeit, in different forms.  An important assumption in the
development of network-delay tomography is that the end-to-end delays
of the individual packets can be measured. This requires that the
sources and the destinations be suitably time synchronized. This is
clearly not an easy task. In fact, the packet-pair techniques were
developed to avoid this problem.  Furthermore, much of the
delay-tomography results were developed assuming that the probes are
multicast packets.  This means that when there are links in series,
the estimation is for the delay across the series and not on the
individual links.

Finally, we remark that the primary objective in this paper is to
develop an analytical framework for `capacity tomogrpahy' and not
develop a practical tool.


\begin{thebibliography}{10}

\bibitem{Vardi96}
Y.~Vardi,
\newblock ``Network tomography: Estimating source-destination traffic
  intensities from link data,''
\newblock {\em Journal of the American Statistical Association}, vol. 91, pp.
  365--377, March 1996.

\bibitem{Abramsson98}
T.~Abramsson,
\newblock ``Estimation of origin-destinaion matrices using traffic counts---a
  literature survey,''
\newblock Tech. {R}ep. IR-98-021, International Insitute of Applied Systems
  Analysis, May 1998.

\bibitem{Zhang03a}
Y.~Zhang, M.~Roughan, C.~Lund, and D.~Donoho,
\newblock ``An information-theoretic approach to traffic matrix estimation,''
\newblock in {\em Proceedings of ACM SIGCOMM}, Kalsruhe, Germany, August 25--29
  2003.

\bibitem{Zhang03b}
Y.~Zhang, M.~Roughan, N.~Duffield, and A.~Greenberg,
\newblock ``Fast accurate computation of large-scale ip traffic matrices from
  link loads,''
\newblock in {\em Proceedings of ACM SIGMETRICS}, 2003.

\bibitem{Medina03}
A.~Medina, N.~Taft, K.~Salamatian, S.~Bhattacharyya, and C.~Diot,
\newblock ``Traffic matrices estimation: Existing techniques and new
  directions,''
\newblock in {\em Proceedings of ACM SIGCOMM}, 2003.

\bibitem{Lakhina04}
A.~Lakhina, K.~Papagiannaki, M.~Crovella, C.~Diot, E.~D. Kolaczyk, and N.~Taft,
\newblock ``Structural analysis of network traffic flows,''
\newblock in {\em Proceedings of ACM SIGMETRICS}, 2004.

\bibitem{Papagiannaki04}
K.~Papagiannaki, N.~Taft, and A.~Lakhina,
\newblock ``A distributed approach to measure ip traffic matrices,''
\newblock in {\em Internet Measurement Conference (IMC) '04}, Taormina, Sicily,
  Italy., October 25--27 2004.

\bibitem{Coates01}
M.~J. Coates and R.~Nowak,
\newblock ``Network tomography for internal delay estimation,''
\newblock in {\em Proceedings of IEEE International Conference on Acoustics,
  Speech, and Signal Processing}, 2001.

\bibitem{Tsang01}
Y.~Tsang, M.~J. Coates, and R.~Nowak,
\newblock ``Passive network tomography using em algorithms,''
\newblock in {\em Proceedings of IEEE International Conference on Acoustics,
  Speech, and Signal Processing}, May 2001.

\bibitem{Castro04}
R.~Castro, M.~J. Coates, G.~Liang, R.~Nowak, and B.~Yu,
\newblock ``Internet tomography: Recent developments,''
\newblock {\em Statistical Science}, vol. 19, no. 3, pp. 499--517, 2004.

\bibitem{Keshav91}
S.~Keshav,
\newblock ``A control-theoretic approach to flow control,''
\newblock {\em ACM SIGCOMM Computer Communication Review}, vol. 21, no. 4, pp.
  3--15, September 1991.

\bibitem{Jacobson97}
V.~Jacobson,
\newblock ``Pathchar: A tool to infer characteristics of internet paths,''
  April 1997.

\bibitem{Downey99}
A.~B. Downey,
\newblock ``Using {pathchar} to estimate internet link characteristics,''
\newblock in {\em Proceedings of ACM SIGCOMM}, September 1999, pp. 222--223.

\bibitem{Jain02}
M.~Jain and C.~Dovrolis,
\newblock ``End-to-end available bandwidth: Measurement methodology, dynamics,
  and relation with tcp throughput,''
\newblock in {\em Proceedings of ACM SIGCOMM}, August 2002, pp. 295--308.

\bibitem{Rebeiro03}
V.~J. Ribeiro, R.~H. Riedi, R.~G. Baraniuk, J.~Navratil, and L.~Cottrell,
\newblock ``{pathChirp}: Efficient available bandwidth estimation for network
  paths,''
\newblock in {\em Proceedings of Passive and Active Measurement Workshop},
  2003.

\bibitem{Prasad03}
R.~Prasad, C.~Dovrolis, M.~Murray, and {k.~claffy},
\newblock ``Bandwidth estimation: metrics, measurement techniques, and
  tools,,''
\newblock {\em IEEE Network}, vol. 17, no. 6, pp. 27--35, November--December
  2003.

\bibitem{Strauss03}
J.~Strauss, D.~Katabi, and F.~Kaashoek,
\newblock ``A measurement study of available bandwidth estimation tools,''
\newblock in {\em Proceedings of the 3rd ACM SIGCOMM conference on Internet
  measurement}, Miami Beach, FL, USA, 2003, pp. 39--44.

\bibitem{Chakraborty05}
S.~Chakraborty, B.~Walia, and D.~Manjunath,
\newblock ``Non cooperative path characterisation using packet spacing
  techniques,''
\newblock in {\em Proceedings of IEEE Conference on High Performance Switching
  and Routing (HPSR)}, Hong Kong PRChina, May 2005.

\bibitem{Dovrolis01}
C.~Dovrolis, P.~Ramanathan, and D.~Moore,
\newblock ``What do packet dispersion techniques measure?,''
\newblock in {\em Proceedings of the IEEE INFOCOM}, Anchorage, AK, USA, April
  2001, pp. 905--914.

\bibitem{Pasztor03}
A.~Pasztor,
\newblock {\em Accurate Active Measurement in the Internet and its
  Application},
\newblock Ph.D. thesis, Department of Electrical and Electronic Engineering,
  The University of Melbourne, February 2003.

\bibitem{Liu05}
X.~Liu, K.~Ravindran, and D.~Loguinov,
\newblock ``What signals do packet-pair dispersions carry?,''
\newblock in {\em Proceedings of the IEEE INFOCOM}, March 2005, pp. 281--292.

\bibitem{Liu07}
X.~Liu, K.~Ravindran, and D.~Loguinov,
\newblock ``A queuing-theoretic foundation of available bandwidth estimation:
  Single-hop analysis,''
\newblock {\em IEEE/ACM Transactions on Networking}, vol. 15, no. 4, pp.
  918--931, August 2007.

\bibitem{Liu08}
X.~Liu, K.~Ravindran, and D.~Loguinov,
\newblock ``A stochastic foundation of available bandwidth estimation:
  Multi-hop analysis,''
\newblock {\em IEEE/ACM Transactions on Networking}, vol. 16, no. 1, pp.
  130--143, February 2008.

\bibitem{Liebeherr07}
J.~Liebeherr, M.~Fidler, and S.~Valaee,
\newblock ``A min-plus system interpretation of bandwidth estimation,''
\newblock in {\em Proceedings of IEEE Infocom}, 6--12 May 2007, pp. 1127--1135.

\bibitem{Park06}
K.~J. Park, H.~Lim, and C.~H. Choi,
\newblock ``Stochastic analysis of packet pair probing for network bandwidth
  estimation,''
\newblock {\em Computer Networks}, vol. 50, pp. 1901--1915, 2006.

\bibitem{Haga06}
P.~Haga, K.~Diriczi, G.~Vattay, and I.~Csabai,
\newblock ``Understanding packet pair dispersion beyond the fluid model: The
  key role of traffic granularity,''
\newblock in {\em Proceedings of IEEE Infocom}, 2006.

\bibitem{Machiraju07}
S.~Machiraju, D.~Veitch, F.~Baccelli, and J.~Bolot,
\newblock ``Adding definition to active probing,''
\newblock {\em ACM Computer Communication Review}, vol. 37, no. 2, pp. 19--28,
  April 2007.

\end{thebibliography}
\end{document}